\documentclass[fleqn,usenatbib]{mnras}

\usepackage{newtxtext,newtxmath}
\usepackage[T1]{fontenc}
\usepackage{ae,aecompl}
\usepackage{soul}
\usepackage{cleveref}
\usepackage{graphicx}	% Including figure files
\usepackage{amsmath}	% Advanced maths commands
\usepackage{amssymb}	% Extra maths symbols

\usepackage{txfonts}
\usepackage{xcolor}
\usepackage{soul}

\usepackage[flushleft]{threeparttable}
\usepackage{subfig}
\usepackage{caption}
\usepackage[export]{adjustbox}

\usepackage{tablefootnote}
\title[1BIGB $\gamma$-ray SEDs]{Extreme \& High Synchrotron Peaked Blazars at the limit of {\it Fermi}-LAT detectability: the $\gamma$-ray spectrum of 1BIGB sources}
 \author[B. Arsioli et al.]{
B. Arsioli$^{1,2,5}$\thanks{E-mail: arsioli@ifi.unicamp.br, bruno.arsioli@ssdc.asi.it},
U. Barres de Almeida$^{3,5}$\thanks{E-mail: ulisses@cbpf.br},
E. Prandini$^{4}$\thanks{E-mail: prandini@pd.infn.it},
B. Fraga$^{3,5}$,
L. Foffano$^{4}$,
\\
$^{1}$Instituto de F\'isica Gleb Wataghin, Universidade Estadual de Campinas (UNICAMP), Rua S\'ergio Buarque de Holanda 777, 13083-859 Campinas, Brazil \\
$^{2}$Science Data Center della Agencia Spaziale Italiana, SSDC - ASI, Rome, Italy \\
$^{3}$Centro Brasileiro de Pesquisas F\'isicas (CBPF), Rua Dr. Xavier Sigaud 150, 22290-180 URCA, Rio de Janeiro, Brasil\\
$^{4}$University of Padova, Department of Physics and Astronomy, and INFN sez. Padova, Italy\\
$^{5}$ICRANet-Rio, CBPF, Rua Dr. Xavier Sigaud 150, 22290-180 URCA, Rio de Janeiro, Brazil       
}

\date{Received: April 23, 2018}

\pubyear{2018}

\begin{document}
\label{firstpage}
\pagerange{\pageref{firstpage}--\pageref{lastpage}}
\maketitle

\begin{abstract}
We present the 1-100 GeV spectral energy distribution for a population of 148 high-synchrotron-peaked blazars (HSPs) recently detected with {\it Fermi}-LAT as part of the First Brazil-ICRANet Gamma-ray Blazar catalogue (1BIGB). Most of the 1BIGB sources do not appear in previous {\it Fermi}-LAT catalogues and their $\gamma$-ray spectral properties are presented here for the first time, representing a significant new extension of the $\gamma$-ray blazar population. Since our sample was originally selected from an excess signal in the 0.3\,-\,500\,GeV band, the sources stand out as promising TeV blazar candidates, potentially in reach of the forthcoming very-high-energy (VHE) $\gamma$-ray observatory, CTA. The flux estimates presented here are derived considering PASS8 data, integrating over more than 9 years of {\it Fermi}-LAT observations. We also review the full broadband fit between 0.3-500\,GeV presented in the original 1BIGB paper for all sources, updating the power-law parameters with currently available {\it Fermi}-LAT dataset. The importance of these sources in the context of VHE population studies with both current instruments and the future CTA is evaluated. To do so, we select a subsample of 1BIGB sources and extrapolate their $\gamma$-ray SEDs to the highest energies, properly accounting for absorption due to the extragalactic background light. We compare those extrapolations to the published CTA sensitivity curves and estimate their detectability by CTA. Two notable sources from our sample, namely 1BIGB~J224910.6-130002 and 1BIGB~J194356.2+211821, are discussed in greater detail. All $\gamma$-ray SEDs, which are shown here for the first time, are made publicly available via the Brazilian Science Data Center (BSDC) service, maintained at CBPF, in Rio de Janeiro.  
%This is a simple template for authors to write new MNRAS papers.The abstract should briefly describe the aims, methods, and main results of the paper.It should be a single paragraph not more than 250 words (200 words for Letters).No references should appear in the abstract.
\end{abstract}


%%%%%%%%%%%%%%%%%%%%%%%%%%%%%%%%%%%%%%%%%%%%%%%%%%

%%%%%%%%%%%%%%%%% BODY OF PAPER %%%%%%%%%%%%%%%%%%

\section{Introduction}

The {\it Fermi}-LAT sources and specially the high-energy catalogues, 2FHL \citep[50\,GeV, ][]{2FHL} and 3FHL \citep[10\,GeV, ][]{3FHL}, provide the best unbiased proxy to the very-high-energy (VHE) extragalactic sky. Meaningful extrapolations can be derived therefrom, as to what are the expectations for future studies in this extreme observational window. 

Complementary to those works is the 1BIGB catalogue~\citep{1BIGB}, which contains 150 extragalactic $\gamma$-ray sources associated to excess signals $\rm > 3 \sigma$ in the 0.3-500 GeV band as observed with {\it Fermi}-LAT. Bearing nearly no overlap with the previous {\it Fermi}-LAT catalogues, the 1BIGB presents sources which were found through targeted analysis of a group of candidate VHE blazars obtained from the 1WHSP and 2WHSP catalogues (\cite{1WHSP} and \cite{2WHSP}), selected on the basis of their IR and X-ray spectral energy distribution (SED) synchrotron properties. Collectively, the objects in the 1BIGB catalogue represent a unique sample from a population of extragalactic objects at the limit of {\it Fermi}-LAT detectability, and which are at the extreme of blazar phenomenology. 

Blazars are active galactic nuclei believed to produce $\gamma$ rays inside relativistic jets which beam the emission towards the observer with bulk Lorentz factors of $\sim 10$ or larger \citep{Urry-Padovani1995}. The $\gamma$-ray production mechanism in blazars is not completely understood, evidence suggesting that most of the emission has a leptonic origin, although a significant but yet unquantified contribution from hadronic processes cannot be excluded \citep{Cerruti2017}. Good sampling of the $\gamma$-ray SED of these objects is therefore fundamental for blazar studies \citep{Bottcher2013}. A related example is the recent effort to unveil new $\gamma$-ray Low-Synchrotron-Peaked (LSP) blazars \citep{LSPdetections}, previously expected to be $\gamma$-ray quiet. 

The SEDs from blazars can extend along many decades in energy, from a few GHz in radio, up to TeV $\gamma$ rays, and are usually characterized by the presence of a synchrotron and an inverse-Compton (IC) bumps in the Log (${\nu})$) vs. Log ($\rm \nu f_{\nu}$) plane \citep{Padovani2017}. The peak-power associated to the synchrotron bump tells us at which frequency most of the AGN electromagnetic power is being released, the parameter Log ($\rm \nu_{peak}^{syn}$) being extensively used to classify blazars. Following discussions from \cite{padgio95}, \cite{BlazarSED} and \cite{ExtremeBlazarsGhisellini1999}, objects with Log($ \rm \nu_{peak}^{syn}$) < 14.5, in between 14.5 to 15.0, and $>$ 15.0 [Hz] have been called, respectively, low-, intermediate-, and high-synchrotron-peak blazars (LSP - ISP - HSP). 

Recent observational data in hard X-rays \citep{2017arXiv171106282C} support the existence of a population of extreme high-energy synchrotron peaked sources (EHSPs). The class had been first proposed by \cite{Costamante01}, being composed of sources with a synchrotron peak frequency above $10^{17}$\,Hz. Clear indications or evidence for the synchrotron peak actually reaching the MeV range is nevertheless still under debate, as in the following works: \cite{2WHSP,Tanaka2014,Kaufmann2011,Tavecchio2011}.

Extreme HSPs are characterized by low luminosity and limited variability, in contrast to typical HSP blazars; the weak variability, however, could be biased by the limited sensitivity of current instruments, rather than to intrinsic characteristics of the sources. The high synchrotron peak (and consequently high IC peak) and low luminosity of EHSPs seem to support the blazar sequence paradigm \citep{1998MNRAS.299..433F, 2017MNRAS.469..255G}. However, the expected shift of both SED peaks to higher energies has not been firmly established yet. We should also call attention to the fact that nearly half of the 1BIGB sources (63 objects) have no measured redshift and another 21 only have an estimated lower-limit redshift. Therefore, as typical of HSP blazars, many 1BIGB sources have no derived luminosity, which could introduce strong bias to the blazar sequence paradigm, as extensively discussed in \cite{1WHSP} with focus to HSP  blazars.

%Since EHSPs are the faintest emitters within the blazar sequence, and are possibly characterized by an IC peak at VHE $\gamma$ rays,
Since EHSPs are possibly characterized by an IC peak at VHE $\gamma$ rays, they are not easy to observe, being hardly detected in current $\gamma$ surveys as well. For example, one of the brightest such sources in X-rays, and the best-studied extreme blazar to date, 1ES~0229+200, was detected by {\it Fermi}-LAT only in the 3FGL, after 4 years of exposure. The HE and VHE $\gamma$-ray spectra for this object were shown to connect smoothly and to be very hard, with no apparent break up to TeV energies~\citep{HESS2007}.

% EHSPs are hard to observe in gamma-rays: given that the Fermi-LAT sensitivity window is not fine-tuned with the IC peak of those sources.

EHSPs draw increasing attention given the possibility that blazars might be associated to astrophysical neutrinos \citep{Padovani_2016_Extreme_blazars_neutrinos,IceCube-NeutrinoTrack-2018}, and ultra-high-energy cosmic rays \citep{Resconi-2017-Blazars-UHECR-Neutrinos}. Extreme blazars have also large impact as probes of the extragalactic background light (EBL) and cosmic magnetic fields \citep{Bonoli2015}. Given the broad context in which blazars play an important role in future astro-particle physics, describing new $\gamma$-ray spectral properties of HSP and EHSP-candidates becomes of major relevance.

An unbiased population study of blazars at VHE $\gamma$ rays is made difficult by the fact that these objects are characterised by intense flaring activity. In fact, many of the known VHE $\gamma$-ray blazars have been detected via triggers based on high states from other bands, rather than through surveys. To produce good estimates on the size and nature of the blazar population potentially detectable at VHEs is therefore an important task. Furthermore, completing the information available on the faint and extreme sector of blazar phenomenology is specially useful, with implications for the determination of their luminosity function (the $\rm \log{N}-\log{S}$ distribution), and the estimates of the total $\gamma$-ray background \citep{Inoue2012,Ackermann2015-EGB}.

%\footnote{An exception would be the 2HWC $\gamma$-ray catalogue \citep{2HWC2017gammacat}, but HAWC's peak sensitivity at 10 TeV greatly hinders its capacity to put meaningful constraints on extragalactic sources.} 

Observations with the current class of VHE instruments, as well as {\it Fermi}-LAT \citep{Tanaka2014}, seem to corroborate with the existence of a population of EHSPs with IC peak above 100 GeV in the $\gamma$-ray band. {\it Fermi}-LAT, for which sensitivity is already greatly decreasing above 100 GeV, might nevertheless be probing only the low-energy side of the IC component, which could easily reach beyond the TeV range. As a result, these putative EHSPs are clearly undersampled in current {\it Fermi}-LAT catalogues, with just three well-studied sources in the local universe up to date -- Mkn 421 and Mkn 501 (at $z \sim 0.03$) and 1ES 0229+200 (at $z$ = 0.14). 

By virtue of its selection methodology, the blazar population in the 1BIGB catalogue is composed of faint HSP blazars and extreme blazars. Until now, a source-by-source discrimination between the two classes has not yet been possible though, due to the limited information available on the synchrotron SED and on the 0.3-500 GeV $\gamma$-ray excess. The fact that most of the objects presented is this work do not show up in previous {\it Fermi}-LAT catalogues, points to a significant new extension of the blazar population, with important implications to the extragalactic science case of currently active VHE $\gamma$-ray observatories, as well as for the future Cherenkov Telescope Array, CTA~\citep{CTA2017arXiv170907997C,AGNunderCTA2013}. 

In this work we analyze the complete sample of 1BIGB sources, using the latest PASS8 {\it Fermi}-LAT data release \citep{PASS8}, and present, for the first time, their 1-100\,GeV $\gamma$-ray spectra. The final list results in the richest high-energy spectral sample of faint HSPs and candidate extreme blazars to date, providing useful new information for the classification of these objects, given that only low-energy SED information was previously available. As we discuss in following \Cref{section:fifth}, the catalogue constitutes also a potential list of targets for current and future VHE observational programmes.

The paper is organised as follows. In \Cref{section:second}, we describe in detail the 1BIGB sample and calculate updated 0.3-100\,GeV broadband fitting for all sources, integrating over 9.3 yrs of currently available data. In \Cref{section:third} we present for the first time, the 1-100\,GeV $\gamma$-ray SEDs of its 148 sources. In \Cref{section:fourth} we stack the $\gamma$-ray SEDs, providing a common view of the source population, discuss about the {\it Fermi}-LAT sensitivity limit at 10 GeV, and present two notable sources from our sample. Finally, in \Cref{section:fifth} we briefly compare our results to the published sensitivity curves of MAGIC, H.E.S.S., and CTA, in order to estimate the detectability and the impact of our source population to current and future VHE $\gamma$-ray studies.  

%In Section 3, we stack the gamma-ray SEDs, normalizing them via the Log($\nu F_{\nu}$) synchrotron-peak parameter, providing a common view of the source population.
%ExtremeBlazars2014: [arxiv.1404.3727]  
%ExtremeBlazarsGhisellini1999:[arxiv.9812202];
%[arxiv.1601.06550]

%%%%%%%%%%%%%%%%%%%%%%%%%%%%%%%%%%%%%%%%%%%%%%%%%%%%%%%%

\section{The 1BIGB sample description}
\label{section:second}

The 1BIGB sample was built based on the selection of 400 $\gamma$-ray candidates from the 1WHSP and 2WHSP samples. Those candidates comprise HSPs with the brightest synchrotron peaks Log\,($\rm \nu f_{\nu}$)\,>\,-12.1 erg/cm$^2$/s, but having no counterpart in previous {\it Fermi}-LAT catalogues -- 1FGL, 2FGL and 3FGL \citep{1FGL,2FGL,3FGL}. A likelihood analysis in the 0.3-500\,GeV band, integrating over 7.2 years of {\it Fermi}-LAT observations, and using the PASS8 data release, unveiled a total of 150 $\gamma$-ray signatures, of which 85 had high-significance, with Test Statistics \citep[TS,][]{Mattox1996} ranging from 25 up to 130 (all >$\rm 5 \sigma$ detections) and the remaining 65 had lower-significance, with TS in between 10 to 25 ($\rm 3 \ to \ 5 \sigma$ detections). Many low and high-significance detections were studied with the help of $\gamma$-ray TS maps, showing a typical point-like source emerging from a smooth background. In addition, the sample showed $\gamma$-ray spectral properties as expected for HSP blazars, with photon spectral index $\rm \langle \Gamma \rangle \sim 1.94$. Given that each of those 150 signatures is associated to a powerful HSP counterpart, they represent robust detections at the edge of $\it Fermi$-LAT sensitivity \citep[as well described by Figure 8 from ][]{1BIGB}, grouped under the acronym 1BIGB.

\subsection{Updating the 1BIGB broadband fit between 0.3-500\,GeV}
\label{update-fit}

The 1BIGB detections considered a broadband likelihood analysis between 0.3-500\,GeV, integrating data from August 04 2008 up to November 04 2015 ($\sim$7.2 years). Since that time, {\it Fermi}-LAT has accumulated more than 2 additional years of observation time (at the time of writing, we have available data up to December 05 2017). Taking advantage of a larger exposure time with {\it Fermi}-LAT, we thus refine the power-law fitting for each 1BIGB source by performing a likelihood analysis similar to \cite{1BIGB}, applying the same setup and data quality cuts. 

%LONG-TABLE-1
\begin{table*}
\caption{Table showing the model description for 13 of the 148 1BIGB $\gamma$-ray signatures, all detected with TS$>$100. Note that a complete table with all the 148 source is available on-line. The first three columns show respectively the 1BIGB source names, right ascension R.A. and declination Dec. in degrees (J2000). The fourth column shows the reported redshifts from literature \citep{bllacz2,pita2013,pks1424p240HighZ,PG1553,shaw2,masetti2013,bll,5BZcat}. The flag ? is used for values reported as uncertain; lower limits are marked with ``$>$" \citep[all lower-limits shown here were derived in][]{1WHSP,2WHSP}, and sources with currently absent redshift were given 0. value. The $\gamma$-ray model parameters from the {\it Fermi} Science Tools assume a power-law to describe the spectrum within the studied energy range 0.3-500\,GeV. The parameter $\rm N_0$ (see \Cref{powerlaw}) is given in units of ph/cm$^{2}$/s/MeV, and $\Gamma$ is the spectral photon index, which are direct outputs from the likelihood analysis over 9.3 years of {\it Fermi}-LAT data in the 0.3-500 GeV band; those results consider the pivot energy fixed as E$_0$\,=\,1\,GeV. The column Flux gives the photon counts in units of ph/cm$^2$/s calculated by integrating \Cref{powerlaw} along the energy range 1-100 GeV, column E-Flux corresponds to the energy flux in units of MeV/cm$^2$/s. For the columns Flux and E-Flux, upper and lower case values represent positive and negative errors, respectively.}
\label{table:1BIGB-9yrs}
{\def\arraystretch{1.5}
 \begin{tabular}{llrcccccc}
\hline
1BIGB Source name  & R.A. (deg)  &  Dec. (deg)  &  z  &  $\Gamma$ & N$_0$ (10$^{-15}$)  & TS & Flux$_{1-100\,GeV}^{( \times 10^{-10})}$ & E-Flux$_{1-100\,GeV}^{( \times 10^{-13})}$ \\
 \hline
  1BIGBJ021631.9+231449  &   34.13333   &   23.24722   &     0.288   &  1.88$\pm$0.10  &   2.49$\pm$0.52   &  116.4 &  2.76$^{ 0.98}_{-0.79}$   & 15.0$^{+4.7}_{-3.3}$ \\ 
  1BIGBJ044240.6+614039  &   70.66917   &   61.6775    &     0.      &  2.01$\pm$0.09  &   4.00$\pm$0.71   &  145.0 &  3.92$^{+1.12}_{-0.95}$   & 17.9$^{+4.3}_{-3.2}$ \\ 
  1BIGBJ050335.3-111506  &   75.8975    &   -11.25167  &  $>$0.57    &  1.85$\pm$0.11  &   2.30$\pm$0.53   &  102.6 &  2.63$^{+1.03}_{-0.82}$   & 15.0$^{+5.1}_{-3.5}$ \\ 
  1BIGBJ050727.1-334635  &   76.86333   &   -33.77639  &     0.      &  1.79$\pm$0.11  &   1.84$\pm$0.43   &  125.0 &  2.25$^{+0.93}_{-0.72}$   & 14.1$^{+5.0}_{-3.5}$ \\ 
  1BIGBJ075936.1+132116  &   119.90042  &   13.35472   &     0.      &  1.75$\pm$0.09  &   2.12$\pm$0.44   &  152.6 &  2.71$^{+0.98}_{-0.79}$   & 17.9$^{+5.5}_{-4.0}$ \\ 
  1BIGBJ082904.7+175415  &   127.27     &   17.90417   &     0.089   &  2.25$\pm$0.10  &   4.47$\pm$0.55   &  136.8 &  3.55$^{+0.79}_{-0.67}$   & 12.1$^{+2.7}_{-2.0}$ \\ 
  1BIGBJ113755.6-171041  &   174.48167  &   -17.17833  &     0.6     &  1.71$\pm$0.10  &   1.90$\pm$0.43   &  126.9 &  2.55$^{+1.00}_{-0.79}$   & 18.0$^{+6.0}_{-4.2}$ \\ 
  1BIGBJ121510.9+073203  &   183.79542  &   7.53444    &     0.137   &  1.64$\pm$0.11  &   1.26$\pm$0.37   &  105.5 &  1.87$^{+0.97}_{-0.72}$   & 14.9$^{+6.2}_{-4.1}$ \\ 
  1BIGBJ144236.4-462300  &   220.65167  &   -46.38361  &     0.103   &  1.92$\pm$0.10  &   3.14$\pm$0.64   &  107.7 &  3.33$^{+1.14}_{-0.93}$   & 17.1$^{+5.0}_{-3.6}$ \\ 
  1BIGBJ154202.9-291509  &   235.5125   &   -29.2525   &     0.      &  1.78$\pm$0.08  &   2.62$\pm$0.50   &  143.2 &  3.27$^{+1.04}_{-0.86}$   & 20.8$^{+5.4}_{-4.1}$ \\ 
  1BIGBJ194356.2+211821  &   295.98417  &   21.30611   &     0.      &  1.44$\pm$0.08  &   2.01$\pm$0.57   &  194.7 &  3.93$^{+1.84}_{-1.44}$   & 42.9$^{+13.2}_{-9.8}$ \\   
  1BIGBJ200204.0-573644  &   300.51708  &   -57.6125   &     0.      &  2.08$\pm$0.10  &   2.93$\pm$0.49   &  105.5 &  2.69$^{+0.78}_{-0.64}$   & 11.2$^{+3.0}_{-2.2}$ \\ 
  1BIGBJ224910.6-130002  &   342.29458  &   -13.00056  &  $>$0.5     &  2.33$\pm$0.01  &  87.01$\pm$1.43   &11240.5 &  64.8$^{+1.93}_{-1.88}$   &202.8$^{+6.2}_{-6.0}$ \\ 
 \hline
 \end{tabular}
 }
\end{table*}
%[Describing the analysis]

We update the 1BIGB power-law parameters and re-evaluate the TS values considering the current 9.3 years of observation, aiming to deliver the most significant description up to date. For this broadband analysis we integrate over the entire 0.3-500\,GeV energy band, using PASS8 data release, and assuming that the $\gamma$-ray spectrum of each $\gamma$-ray source can be described by a power-law:

\begin{equation} \rm
%\vspace{5pt}
%$
\hspace{85pt}\frac{dN}{dE} \, \text{=} \, N_0 \; \left( \frac{E}{E_0} \right)^{-\Gamma} \;,
\label{powerlaw}
%$
\end{equation} 
\vspace{4pt}

\noindent where $\rm E_0$ is a scale parameter (also known as pivot-energy), $\rm N_0$ is the prefactor (normalization) corresponding to the flux density in units of ph/cm$^{2}$/s/MeV at the pivot energy $\rm E_0$, and $\rm \Gamma$ is the photon spectral index for the energy range considered. Both $\rm \Gamma$ and $\rm N_0$ are set as free parameters and further adjusted by the likelihood fitting routine. Source positions and E$_0$\,=\,1000\,MeV are set as fixed  parameters, kept constants for the analysis. In the source-input xml file, all sources within 10$^\circ$ from the candidate had both $\rm \Gamma$ and $\rm N_0$ parameters flagged as free\footnote{In this regard we are following recommendations from Fermi Science Tools user guide \url{https://fermi.gsfc.nasa.gov/ssc/data/analysis/scitools/binned_likelihood_tutorial.html} which advise users to set free parameters at least within 7$^\circ$ from the source of interest. This is a consequence of the large point spread function (PSF) specially at the low-energy threshold, which can overlap with nearby sources. Therefore, in order to get a confident description of a particular source, we need to properly fit and adjust the whole environment around it.}, and therefore their 3FGL models that are based on 4 years of observations will be adjusted. This particular choice increases the computational burden of the analysis, but is crucial for adapting the model maps to the extra 5.3 years of exposure being considered. A sub-sample of the results, listing only 13 cases detected with TS$>$100 is shown in \Cref{table:1BIGB-9yrs}; a list with the total 148 cases which had good convergence will be available as on-line material. %, Sec \ref{appendix1}.  

\begin{figure}
\includegraphics[width=1.0\linewidth]{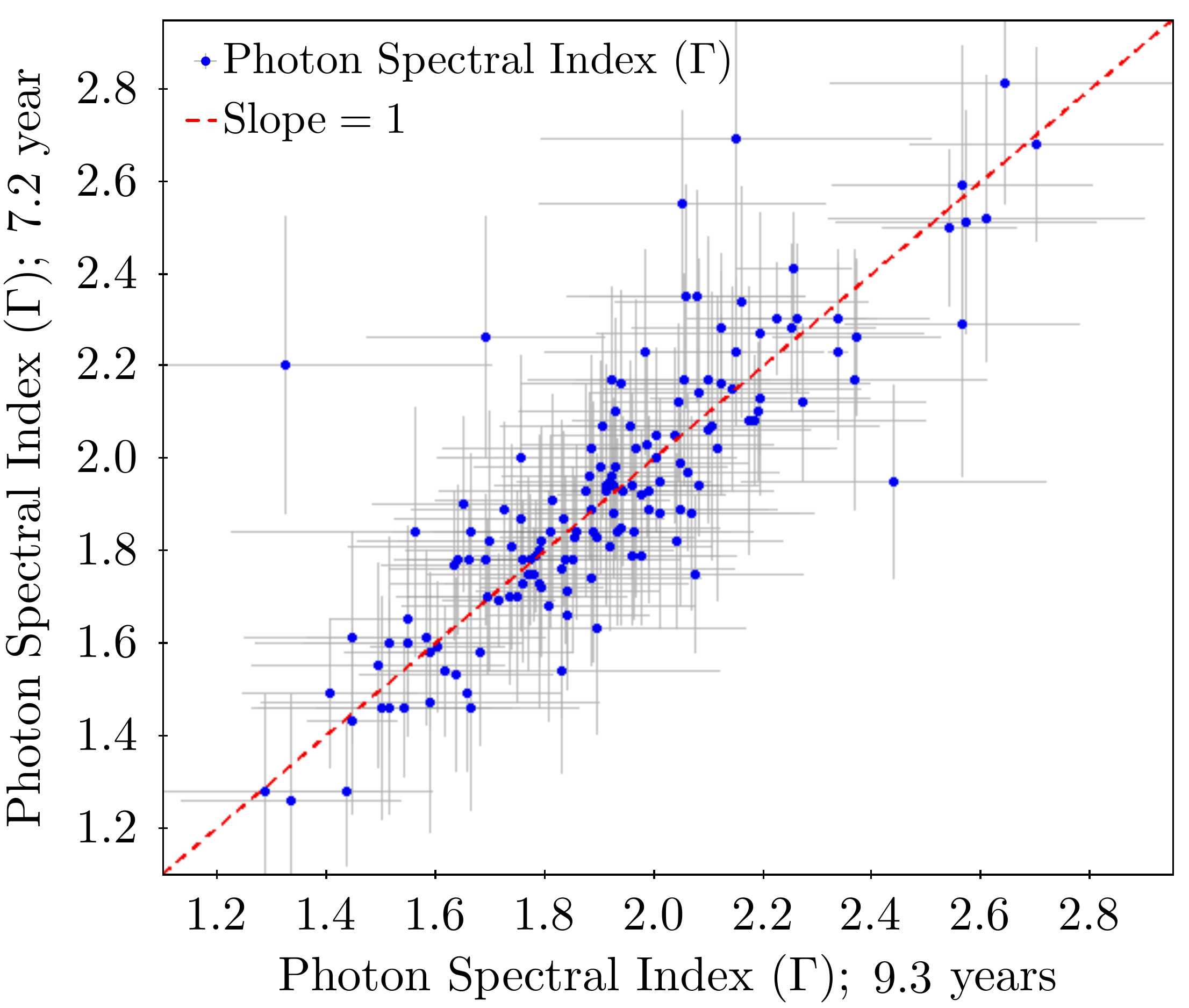}
\caption{Comparison between the photon spectral index ($\Gamma$) integrating over for 7.2 years of exposure, with the updated values calculated with 9.3 years of observations with {\it Fermi}-LAT. The $\Gamma$ factor considers a power-law fit to the entire 0.3-500 GeV energy band, and its corresponding error bars are represented in gray. The red-dashed line is a curve with slope equal one.}
\label{fig:VarIndex}
\end{figure}

For most cases the $\Gamma$ parameter estimates obtained by integrating over 7.2 ($\Gamma_{7.2}$) and 9.3 ($\Gamma_{9.3}$) years are relatively stable, since the intrinsic variability is well contained within the error bars (compared to the red-dashed line in \Cref{fig:VarIndex}, for $\rm \Gamma_{7.2}$\,=\,$\rm \Gamma_{9.3}$). There were only two cases which did not show good likelihood-convergence when integrating along 9.3 yrs, namely 1BIGB~J080135.8+463824 and 1BIGB~J145508.2+192014, which were removed from the 1BIGB SED list. We would like to note that those sources had TS close to 10 in the previous 1BIGB sample (7.2 years). Few non-detections are expected when integrating over different time-windows, specially owing to the uncertainty induced by source variability.

\section{The $\gamma$-ray spectral energy distribution}
\label{section:third}
The 1BIGB sample is composed of $\gamma$-ray sources which are, on average, faint and close to the detectability limit of {\it Fermi}-LAT. In the process of building the $\gamma$-ray SEDs, we had to deal with the problem of poor photon counts when integrating over a short energy bin. In most cases, simply dividing the broadband energy (between 1\,GeV and 500\,GeV) in equally spaced logarithmic bins result in low TS detections per bin, which most of the times only become upper limits for the SED. This is the usual way SEDs are built, and inevitably a considerable amount of relevant information available via broadband analysis is not being incorporated. 

%[Br] [THE 1GeV cut is explained just after] [I think it is better to explain first why we need to use overlapping energy bins. ]

In order to capture the shape of the $\gamma$-ray spectrum and estimate reliable SED data points, we integrate each energy bin over a large energy band (larger than `equally spaced logarithmic bins') and evaluate the flux at a specific energy for each bin. In other words, we take into account information from a broadband analysis by accepting superposed energy channels. 

Table \ref{E-bin} describes the energy bands over which we integrate in order to estimate fluxes at specific pivot-energies ($\rm E_0$), later used to build the $\gamma$-ray SEDs. We choose those E$_0$ values to be close to equally spaced bins in Log(E) scale, with increments of the order of $\sim$10$^{0.25}$. For example, starting from the 1\,GeV SED point we integrate over 1.0 to 5.0 GeV, adjusting a power-law (see \Cref{powerlaw}) to this interval and setting E$_0$\,=\,1\,GeV, so that N$_0$ represents the flux at 1\,GeV when taking into account the spectral trend in the selected energy-bin. For the later SED point at 1.7\,GeV, we increment the energy band by a factor $\sim$10$^{0.25}$, therefore integrating over 1.0 to 10\,GeV, so to capture the power-law trend when estimating an SED point at higher energy. We proceed in this way up to the 5\,GeV SED point, and for the next ones we start incrementing both the low and high-energy band by a factor $\sim$10$^{0.25}$, till we reach the 100\,GeV SED data point, with a high-energy threshold of 500\,GeV.

\begin{table}
	\centering
	\caption{Definition of energy bins used for the broadband analysis, to estimate fluxes N$_0$ at each pivot energy $\rm E_0$. The estimate is done via likelihood analysis, adjusting a power-law to each energy bin.}
	\label{E-bin}
	\begin{tabular}{l|l} % four columns, alignment for each
		\hline
		E$_0$ [GeV] & Integrate over [GeV] \\
		\hline
		1.0   & 1.0 - 5.0     \\
		1.7   & 1.0 - 10.0     \\
		3.0   & 1.0 - 17.0     \\
        5.0   & 1.0 - 30.0     \\
		10.0  & 1.7 - 50.0     \\
		17.0  & 3.0 - 100.0    \\
        30.0  & 5.0 - 170.0    \\
		50.0  & 10.0  - 300.0  \\
		100.0 & 30.0  - 500.0  \\
		\hline
	\end{tabular}
\end{table}

%1 GeV flux   : Integrate over 1   - 5 GeV
%1.7 GeV flux : Integrate over 1   - 10 GeV
%3 GeV flux   : Integrate over 1   - 17 GeV
%5 GeV flux.  : Integrate over 1   - 30 GeV
%10 GeV flux  : Integrate over 1.7 - 50 GeV
%17 GeV flux. : Integrate over 3  - 100 GeV
%30 GeV flux. : Integrate over 5  - 170 GeV
%50 GeV flux  : Integrate over 10 - 300 GeV 
%100 GeV flux.: Integrate over 30 - 500 GeV 

Considering the total of 1350 likelihood analysis (integrating over 9.3 yrs) needed to be computed for building the entire SED catalogue, we limit the low energy band to 1 GeV to improve the computation time. Given that the photon counts at 1\,GeV and higher energy levels are relatively reduced with respect to lower energy bands, working with a cut at 1\,GeV allows the likelihood to converge in a reasonable time (so that a large scale analysis could be performed). The cut in energy also adjusts well to the hard $\gamma$-ray spectral slope expected for HSP sources; taking into account that the detector PSF improves with increasing energy, the energy cut also helps to avoid contamination from nearby sources. 

In the source-input xml file, all sources within 10$^\circ$ from the candidate are set free to vary their spectral fitting parameters, just as used in previous broadband analysis (\Cref{update-fit}). Both $\Gamma$ and $\rm N_0$ are set as free parameters and further adjusted by the \texttt{gtlike} fitting routine. The source position and the scaling $\rm E_0$ are set as fixed parameters, with $\rm E_0$ changing only according to the SED point being calculated. At this point, we call attention to the choice of the pivot-energy $\rm E_0$, which corresponds to the energy where the differential flux $\rm N_0$ [ph/MeV/cm$^2$/s] is going to be estimated. Later, we compute the flux [erg/cm$^2$/s] at a given $\rm E_0$ simply converting\footnote{The factor 1.602$\times$10$^{-6}$ is used to convert MeV to erg.}: $\text{flux = }\rm N_0 \cdot E_0^2 \cdot 1.602 \times 10^{-6}$, and cases showing TS\,$<$\,6.0 per bin are given upper limits. This is how we combine a broadband analysis with the flux calculation for each energy bin, to deal specifically with faint sources. 

For the highest energy channels 50\,GeV and 100\,GeV, a broadband binned analysis suffers from rather low photon counts, to which an unbinned analysis is much better adapted. Therefore, for those two channels we have applied unbinned analyses, assuring a good SED agreement between the few 1BIGB sources which are also part of the 2FHL and 3FHL catalogues. As an example\footnote{Please, consult \cite{1BIGB} for a complete list of 1BIGB sources having 2FHL counterparts} we cite the cases of 1BIGB~J044240.6+614039, 1BIGB~J030433.9-005403 and 1BIGB~J113046.0-313807. %shown in Fig. \ref{fig:bigb_hess}. %cases the 50\,GeV flux as measured from other catalogue for the few cases where the 3FHL and 1BIGB . 

%%%%%%%%%%%%%%%%%%%%%%%%%%%%%%%%%%%%%%%%%%%%%%%%%%%%%%%%
  
\section{Average $\gamma$-ray behavior and blazar populations}
\label{section:fourth}
\begin{figure}
\centering
\includegraphics[width=1.0\linewidth]{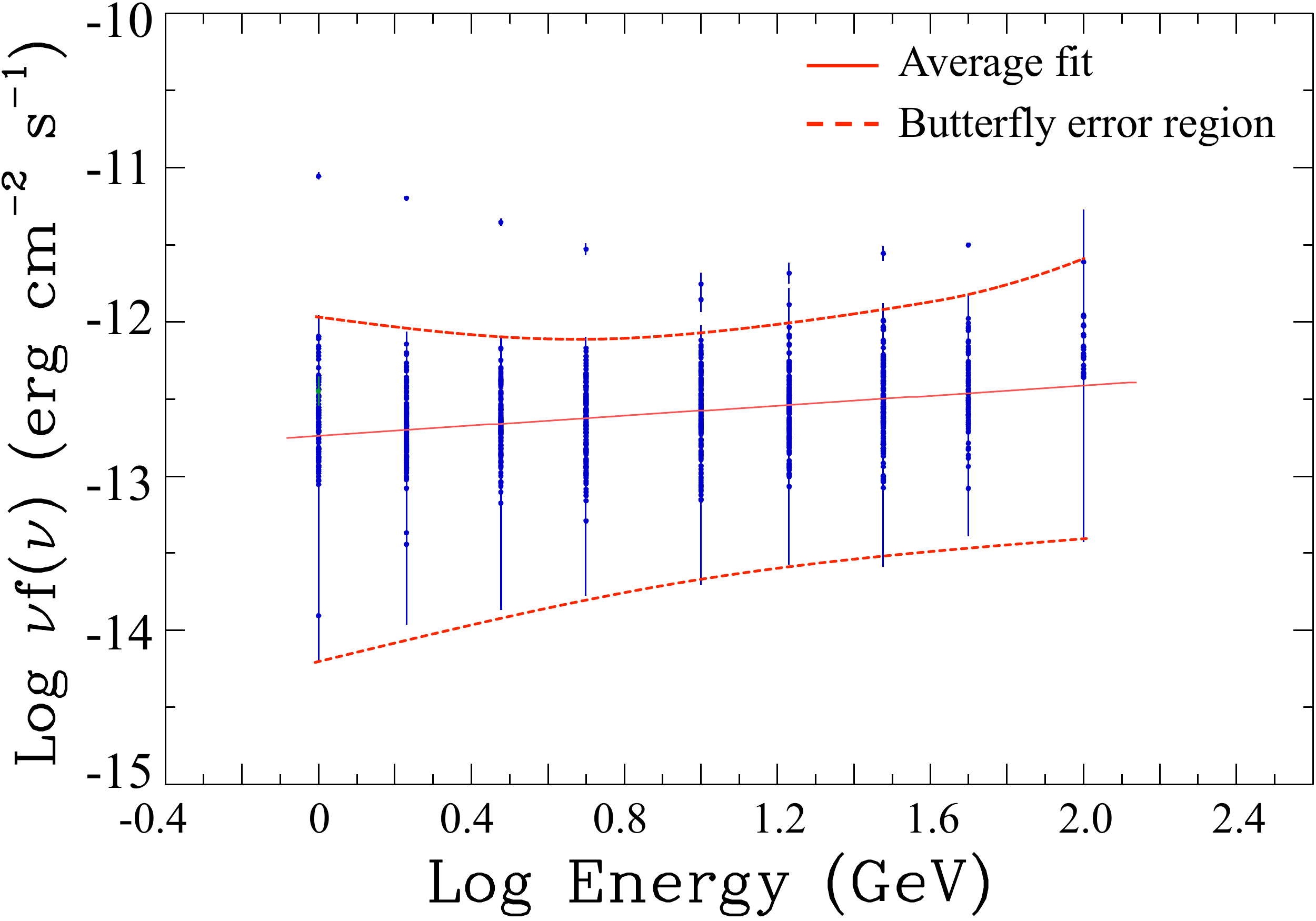}
\caption{Stacking of the 148 1BIGB SEDs in the 1-100\,GeV energy band. Red dashed lines enclose the main flux density interval covered by the measured fluxes and its uncertainties, showing the main region in Log ($\rm \nu$) [GeV] vs. Log ($\rm \nu f_{\nu}$) [eg/cm$^2$/s] covered by the new $\gamma$-ray spectrum. Red line is a average power-law fit considering all the 148 SEDs.}
\label{fig:stack1}
\end{figure}

The 1BIGB catalogue contains average spectral information about the sources derived from a 0.3-500\,GeV broadband analysis. With the purpose of investigating in more detail the spectral properties of these faint objects, here we present the 1-100\,GeV spectral points resulting from the analysis of 9.3 years of {\it Fermi}-LAT data as reported in the previous Section. As can be seen in \Cref{fig:stack1}, the spectral points populate the full energy window under consideration, except for the highest energy bin, around 100\,GeV, where only 13\% of the blazars present a significant signature.% (most cases only allowed an upper limit to be calculated).

When considering an individual power-law fit to the $\gamma$-ray spectrum, the derived photon spectral index covers a wide range from 1.26 to 2.81, with an average error on the slope of the order of 10\%. The large spread in spectral slopes suggests that the objects composing this catalogue do not present any common spectral feature at the HE $\gamma$-ray band. No characteristic spectral turnover, nor spectral break, correlating with other  properties such as synchrotron-peak frequency or flux level was found.% as well.

At first glance such results may look at odds with expectations from a description following the blazar sequence \citep{EHSP-2001-Costamante}; we remind, nevertheless, the fact that an effective population study is hardly possible here, given the limited sample size and the lack of homogeneity in the sample selection, which is composed of objects distinguished only by being at the very limit of {\it Fermi}-LAT detectability. The 1BIGB source population is therefore likely composed of a mixture of faint HSPs and extreme-HSP blazars.

%\br{[1BIGBJ225147.5-320611 , this source is a good example of SED which has a break. Broadband fitting gives $\Gamma$=2.07. But fitting only E>1GeV data points give $\Gamma$=1.04 . Maybe we should we describe this better along the text. ]}

Within the population, two sources stand out showing a $\gamma$-ray spectrum which clearly exceed most objects in flux. All the remaining sources are clustered in a region of $\nu F_{\nu}$ between 10$^{-13}$ and 10$^{-12}$\,erg/cm$^2$/s, with no clear common spectral trend correlated to synchrotron peak information. The two peculiar sources will be further discussed in \Cref{subsec:NotableSources}. 
%In the spectral region under consideration, between 1-100\,GeV,

We fitted the clustered points for the remaining sources with a power-law (red line in \Cref{fig:stack1}) and found an average photon spectrum slope of $\Gamma \sim $1.86. This suggests that for the 1BIGB population as a whole, the peak of the second SED bump is located at energies larger than those covered by our analysis, falling within the VHE $\gamma$-ray range. This result is consistent with expectations for EHSPs, which should have a hard spectrum at high-energies, such as those measured for 1ES~0229+200 and 1ES~1011+496 \citep{HESS2007, 2016AA...590A..24A}, the two hardest spectrum sources ever detected in the VHEs. The large spread in spectral points and fit slopes, however, make these considerations on the average trend not representative of all the sample, which appears non-uniform except for the faintness in flux.

%Now, normalizing based on the Log($\nu$f$_{\nu})$ parameter. The mean value $\langle$Log($\nu$f$_{\nu})$ $\rangle$=-11.85$\pm$0.02    [$\sigma$ = 0,24].
%[ToDo][Br] Rebuild the catalogue normalizing each flux bin, case by case. Multiply each flux bin by a factor \textit{f}, where:   
%\begin{equation}
%f =  \rm{ 1 -  \frac{\langle Log(\nu f _{\nu}) \rangle - Log(\nu f _{\nu})}{\langle Log(\nu f _{\nu}) %\rangle } }   
%\end{equation}

\begin{table}
	\centering
	\caption{Most extreme 1BIGB sources (highest synchrotron peak frequency). The index is calculated by fitting a power-law to the SED points in the range 1-100 GeV.}
	\label{extreme_sources}
	\begin{tabular}{l|l|l} % four columns, alignment for each
		\hline
		Name  & $\Gamma_{1-100\,GeV}$ & Log $\nu_{peak}^{syn}$\\
		\hline
		1BIGB J194356.2+211821   & $1.68\pm 0.23$ & 18.1    \\
		1BIGB J151618.7-152344   & $2.08\pm 0.07$ & 18.0     \\
		1BIGB J225147.5-320611   & $2.04\pm 0.07$ & $>$18.0    \\
        1BIGB J020412.9-333339   & $1.60\pm 0.08$ & 17.9     \\
		1BIGB J032056.2+042447  & $2.65\pm 0.18$ & 17.9     \\
		1BIGB J050419.5-095631  & $2.21\pm 0.08$ & 17.9    \\
        1BIGB J055716.7-061706  & $1.81\pm 0.05$ & 17.9    \\
		1BIGB J125341.2-393159  & $1.72\pm 0.09$ & 17.9  \\
		1BIGB J132541.8-022809 & $1.83\pm 0.15$ & 17.9  \\
        1BIGB J160519.0+542058 & $2.18\pm 0.14$ & 17.9 \\
		\hline
	\end{tabular}
\end{table}

\subsection{Population considerations} 

%As previously mentioned, 
%{Br]There was no mention before, so I just comment that line.
%{\color{red}[Should we say EHSP - candidates?]}
Based on synchrotron peak values reported in the 2WHSP catalogue, the 1BIGB sample holds 52 EHSP candidates, of which 11 objects do not have a firm determination of the peak. We classify the remaining 96 objects of the 1BIGB as (faint) HSP blazars for the purposes of this study.

%We remove two HSPs ( log(nu_peak ) < 17. 
%There were only two cases which did not show good likelihood-convergence when integrating along 9.3 yrs, namely 1BIGB~J080135.8+463824 and 1BIGB~J145508.2+192014, which were removed from the 1BIGB SED list

In order to investigate the properties of EHSP and HSP populations within the catalogue, we have compared the SEDs of the two groups of sources. We find that both the average slope of the EHSP candidates and their spanned fluxes are comparable to those of the HSPs. Therefore, the two populations have no clear signature in the SEDs analyzed which would allow distinguishing them on the basis of their {\it Fermi}-LAT $\gamma$-ray properties alone. 

This result suggests that there are EHSPs whose IC peak behaves like standard HSPs, and therefore a shift of the synchrotron peak location to extreme energies does not always reflect in a shift of the second SED peak. Variability most certainly plays an important role on the detectability of these faint sources and a bias towards the presence of high states can contaminate the sample mean, affecting also its spectral properties.

For cases where the second peak is shifted, this could be detected in the VHE regime with the current IACTs or the forthcoming CTA. In any case, the sub-class of EHSPs displaying an IC peak location in the VHE $\gamma$-ray range is certainly one of the primary targets of VHE $\gamma$-ray observations, and might represent the only class of blazar sources partially undetected by {\it Fermi}-LAT, but well within the reach of IACTs.

\subsection{Sensitivity limit}  

\begin{figure}
\centering
\includegraphics[width=1.0\linewidth]{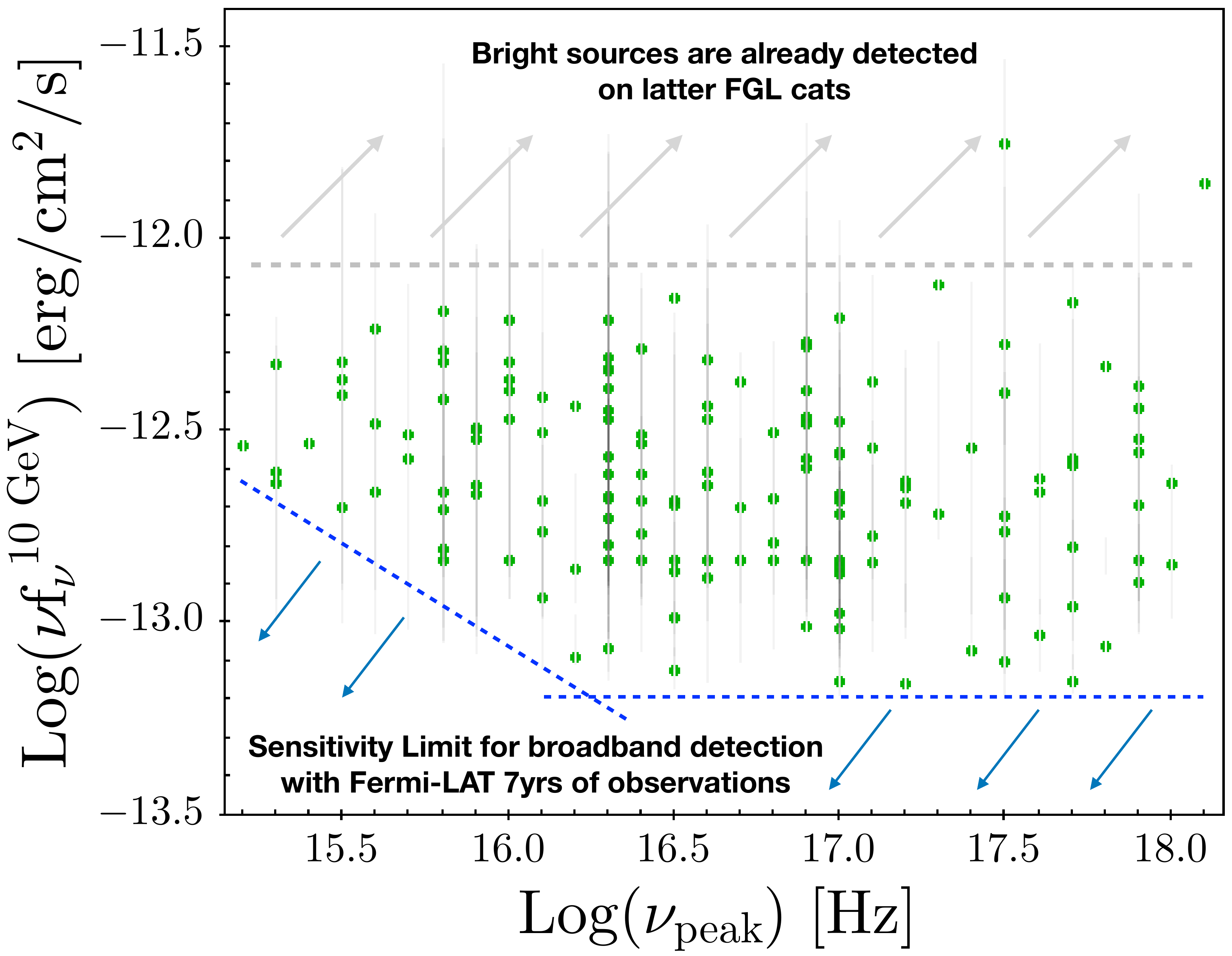}
\caption{Logarithm of flux density (erg/cm$2$/s) at 10\,GeV vs. logarithm of synchrotron peak frequency (Hz) for the 148 1BIGB sources considered in the study. Dashed lines show the previous (gray) and improved (blue) sensitivity limits.}
\label{fig:SensitivityLimit}
\end{figure}

The 1BIGB sources give us the opportunity to study the sensitivity limit for the $\gamma$-ray detection of HSP and EHSP blazars with {\it Fermi}-LAT. \Cref{fig:SensitivityLimit} shows the flux density at 10\,GeV as a function of the synchrotron peak position taken from the 2WHSP catalogue. Except for two peculiar cases, all sources are contained within a region of flux density at 10 GeV comprised between 10$^{-12.1}$\,erg/cm$^2$/s and 10$^{-13.2}$\,erg/cm$^2$/s. The sources presented in the 1BIGB catalogue explore a new range of sensitivity, previously not accessible to {\it Fermi}-LAT catalogs due to the lower exposure time (4 years, in case of the 3FGL). %But also we see a limitation for detecting sources with 10 GeV flux density lower than 10$^{-13.2}$ ergs/cm$^2$/s . There is a clear cut at Log($\nu f _{\nu}$) $\approx$ -12.1 , most of sources brighter than that have already been detected in previous FGL point source catalogs.   
    
The cut at the bottom left corner seems to represent the increasing difficulty for detecting sources with the lowest synchrotron Log ($\nu^{Syn}_{peak}$) values. A possible explanation is that for low synchrotron-peak values, the IC peak may move to lower energies, hindering the detections; but also this could be connected to incompleteness of the 2WHSP sample close to the synchrotron peak cut, at $\nu \sim$10$^{15}$\,Hz. As reported in \cite{2WHSP} the incompleteness might be induced by the multi-frequency selection criteria which was used to build the 2WHSP sample. % based on radio to infra-red, to X-ray slopes. 

%We also notice the lack of statistics at the highest peak frequencies. 
%%%%%%%%%%%%%%%%%%%%%%%%%%%%%%%%%%%%%%%%%%%%%%%%%%%%%%%%

%%\begin{figure}
%\centering
%\includegraphics[angle=0, width=1.0\linewidth]{lucaplots/EHBL_1BIGB_all.png}
%\caption{ TO BE REMOVED {\it Fermi}-LAT spectra of the EHBL sources (each one with a different colour) compared to  the other HSP blazars of the 1BIGB sample (in black).}
%\label{EHBL_subsample}%
%\end{figure}

 \begin{figure*}
\subfloat[][]
{\includegraphics[width=0.485\linewidth]{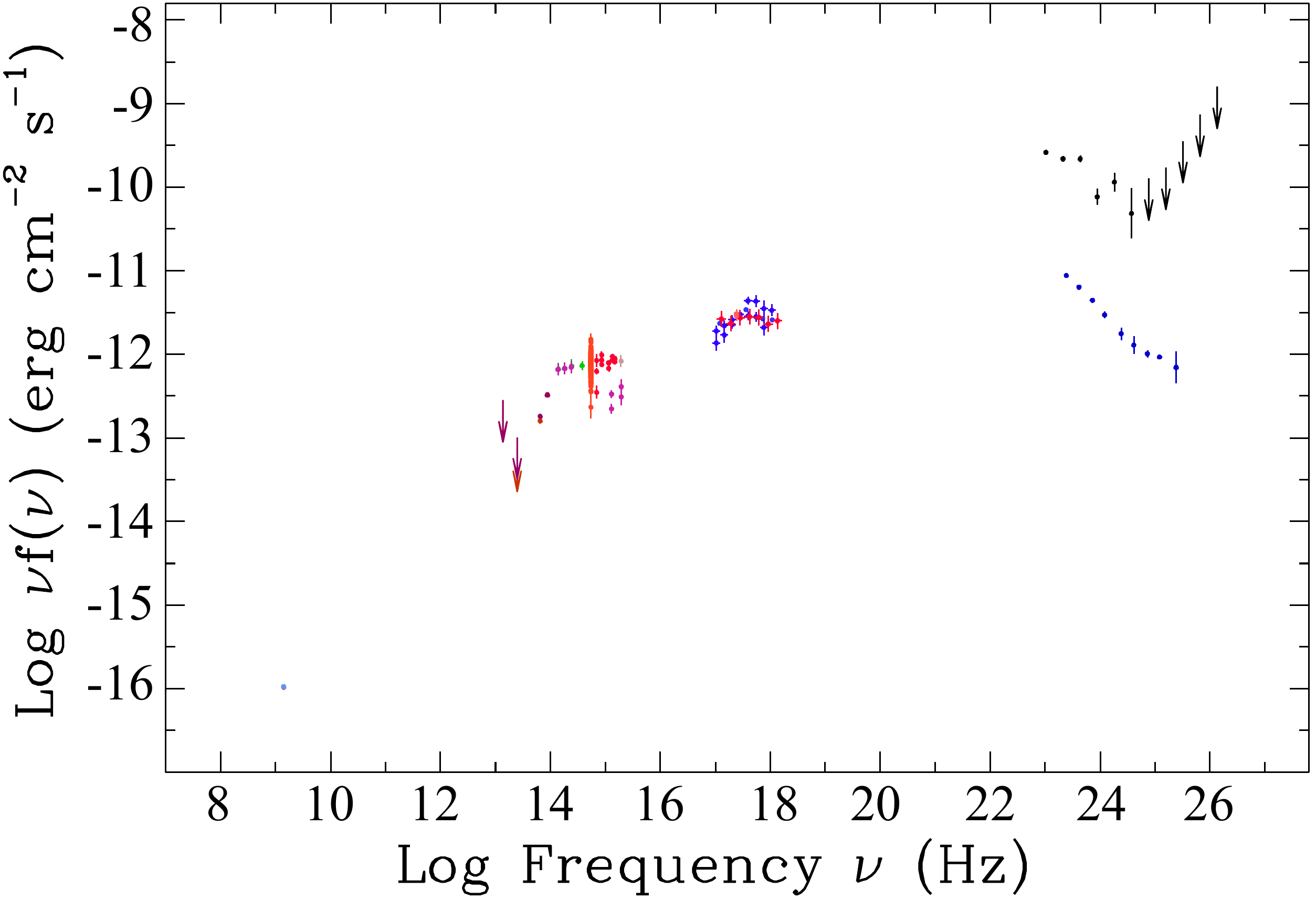} \label{fig:SED1a}}
\subfloat[][]
{\includegraphics[width=0.5\linewidth]{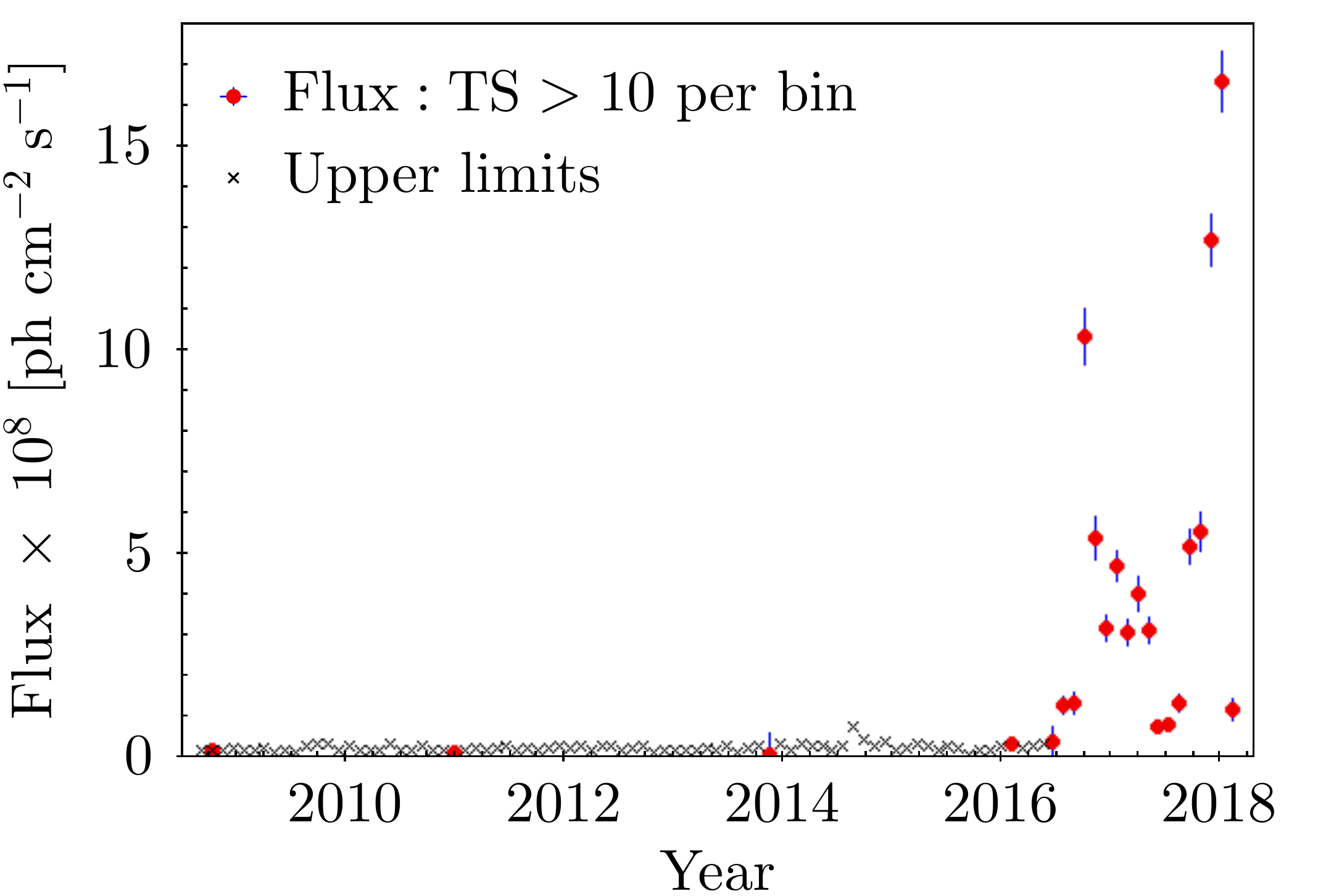} \label{fig:SED1b}}
\caption{Left plot: SED of 1BIGBJ224910.6-130002, showing the mean $\gamma$-ray SED obtained by integrating over 9.3 years in blue, and during the 25 days flare peak (MJD 57662.65-57687.65) in black. Right plot: Light curve of 1BIGBJ224910.6-130002 considering one month time bins, integrating over the broadband of 900\,MeV up to 500\,GeV. Flux calculated only for >\,3$\sigma$ bins (red points), error bars and upper limits (black cross) are calculated using Integral method considering 95\% of confidence level.}
\label{fig:SED1}
\end{figure*}

\subsection{Notable sources}\label{subsec:NotableSources}

%[Br] Keep in mind: Swift proposal to observe $\approx$ 80 1BIGB sources is currently underway. We could recalculate/update Log($\nu$) and Log($\nu$f$_{\nu}$ at synchrotron peak for all Swift-observed sources; And also show the SED for cases where Swift data brings relevant contribution (like resolving new extreme-HSPs).  

The stacked $\gamma$-ray SEDs displayed in \Cref{fig:stack1} present two distinct outliers, namely the sources 1BIGB~J224910.6-130002 and 1BIGB~J194356.2+211821. 

%Interestingly the {\it Fermi}-LAT spectrum obtained with over 9.3 years of data shows a steep decaying behavior ($\Gamma = -1.65$ in $\nu F  \nu$ space) \bf{TO BE CONFIRMED}
%{br] From the 9.3 yrs analysis, the Photon spectral Index \Gamma is 2.33}
The source 1BIGB~J224910.6-130002, also known as RBS~1899, is a blazar of unknown redshift classified as an EHSP due to its synchrotron peak position estimated at 10$^{17.5}$\,Hz. Interestingly the {\it Fermi}-LAT spectrum obtained with over 9.3 years of data shows a steep decaying behavior in Log ($\nu$) versus Log ($\nu F_{\nu}$) space, with photon spectral index $\Gamma \sim 2.33$ and relatively bright in $\gamma$ rays, as seen in \Cref{fig:SED1a}. However this unusual behavior is probably due to the averaging of the quiescent state with a series of $\gamma$-ray flares which happened along 2016 to 2017, as can be seen in the light curve reported in \Cref{fig:SED1b}. During these flaring episodes, the flux increased of $\sim$40 times with respect to the quiescent phase. This may also explain why the source was undetected in previous FGL and FHL catalogues. This is an important example of a $\gamma$-ray blazar which could represent a major target for the future CTA during periods of extreme activity (as detected during MJD 57662 - 57687, \Cref{fig:SED1a} in black). It also illustrates the strong impact of flaring activity in attempting a population study of extreme or faint blazars, since the most extreme $\gamma$-ray activity is smoothed over time when integrating over years of {\it Fermi}-LAT observations. 

\begin{figure}
\includegraphics[width=1.0\linewidth]{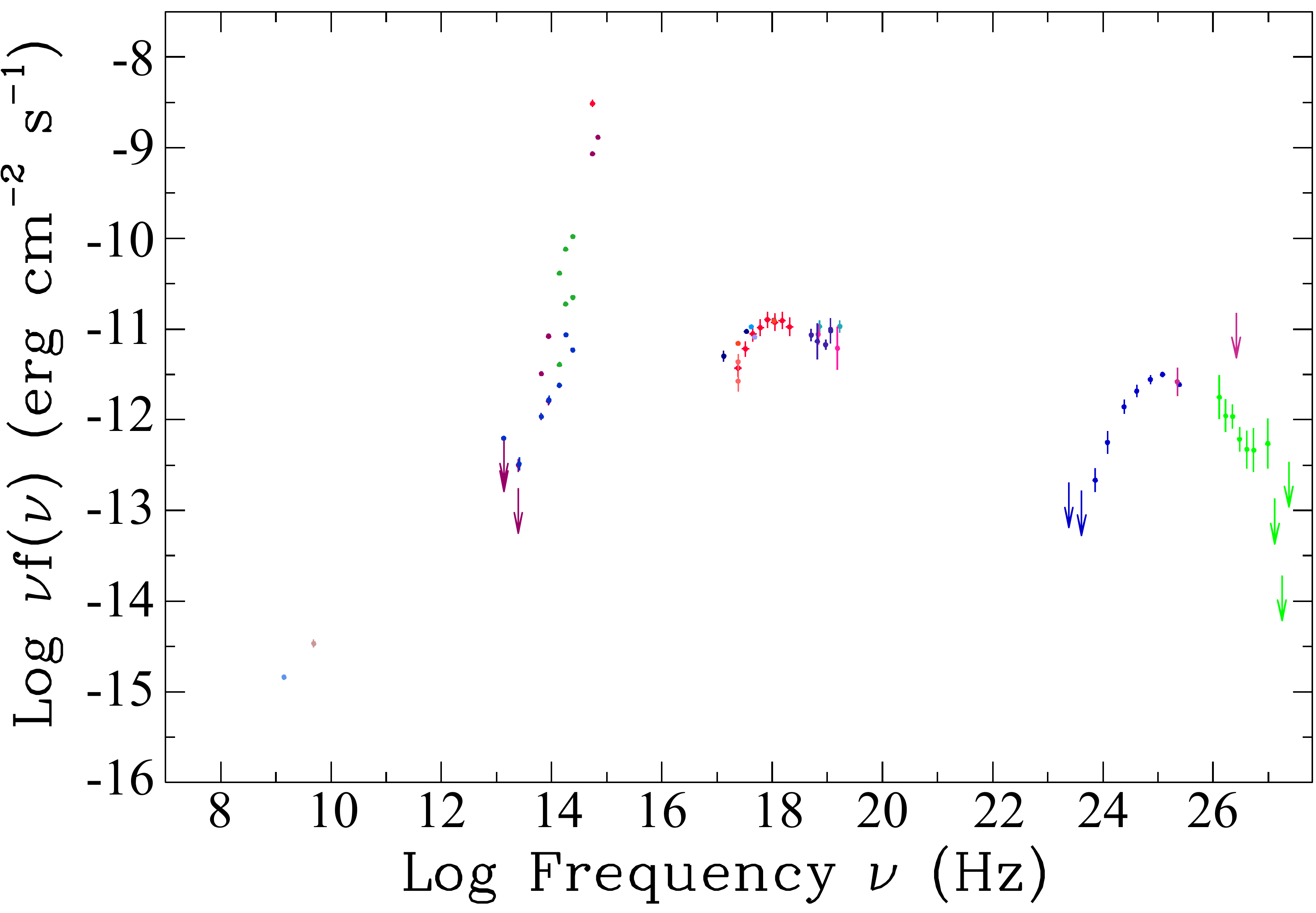}
\caption{SED of 1BIGB~J194356.2+211821. In the $\gamma$-ray band we have data points corresponding to our new {\it Fermi}-LAT detection (in blue), 3FHL (in magenta), and H.E.S.S. (in light green). Notice at 100\,GeV the good agreement between 1BIGB (highest energy point) and 3FHL (lowest energy point).}
\label{fig:bigb_hess}
\end{figure}

The second peculiar source, 1BIGB~J194356.2+211821 also known as HESS~J1943+213, is the most extreme blazar of the 1BIGB catalogue, with Log ($\rm \nu_p$) = 18.1. %(not considering the lower limits). 
It was serendipitously discovered by the H.E.S.S. Collaboration \citep{HESS} at VHE $\gamma$ rays during a galactic plane survey and later confirmed with VERITAS observations \citep{veritas1943}. Due to the source's position on the galactic plane and its multi-wavelength properties, the classification was difficult and three competing hypotheses were proposed: a pulsar wind nebula (PWN), a $\gamma$-ray binary, and a blazar, with the latter being favored. Previous studies tried to analyse \textit{Fermi}-LAT data in order to determine the nature of the source: \citet{fermidetection-Pass7} analysed Pass 7 data on the position of the HESS source, finding a weak detection (TS=22.3) in the 10-300 GeV range, with a rather soft spectral index of $\Gamma\approx 2.4$; analysis of the Pass 7 Reprocessed data \citep{fermidetection-Pass7R} shows a detection near the position determined by HESS with TS=36.0 in the range 1-300 GeV, and a spectral index of $\Gamma$\,=\,1.59 (in agreement with the value reported here considering the errors). Both these studies favor the blazar interpretation based on multiwavelength data and further observations seem to corroborate this (see e.g. \citet{hess1943bllac} and \citet{hess1943bllac-2}), making this the first blazar serendipitously detected by VHE $\gamma$-ray ground-based instruments. Interestingly, the optical and UV properties alone would not be enough to classify this source as a blazar, showing the importance of the multi-wavelength approach. The SED in \Cref{fig:bigb_hess} shows that the VHE $\gamma$-ray data is complemented nicely by our {\it Fermi}-LAT points, making the IC peak location clearly above few hundred GeV. 
%{\bf [EP suggestion: IC peak location above few hundred GeV.]})  [Done]

\begin{table}
\caption{Five sources selected for the VHE $\gamma$-ray extrapolation study.}             % title of Table
\label{tab:sources_cta}      % is used to refer this table in the text
\centering                          % used for centering table
\begin{tabular}{c c c c c}        % centered columns (4 columns)
\hline\hline                 % inserts double horizontal lines
1BIGB\,J Source Name & $ z $ & log $\nu_{p} $ & $\Gamma$  & Cut-off \\    % table heading 
\hline                        % inserts single horizontal line
090802.2$-$095936 & 0.053 & 17.6    & $1.65\pm 0.13$ & 600\,GeV\\
151041.0+333503 & 0.114 & $>$17.5 & $1.92\pm 0.15$ & 300\,GeV \\
225147.5$-$320611 & 0.139 & $>$18    & $2.04\pm 0.07$ & 1\,TeV\\
220155.8$-$170700 & 0.169 & 17.7    & $0.90\pm 0.15$ & 150\,GeV\\
223301.0+133601 & 0.214 & $>$17   & $1.76\pm 0.15$ & 1\,TeV\\
\hline                                   %inserts single line
\end{tabular}
\end{table}

%1BIGB J090802.2-095936 & 0.053 & 17.6    & $1.65\pm 0.13$ & 600\,GeV\\
%1BIGB J151041.0+333503 & 0.114 & $>$17.5 & $1.92\pm 0.15$ & 300\,GeV \\
%1BIGB J225147.5-320611 & 0.139 & $>$18    & $2.04\pm 0.07$ & 1\,TeV\\
%1BIGB J220155.8-170700 & 0.169 & 17.7    & $0.90\pm 0.15$ & 150\,GeV\\
%1BIGB J223301.0+133601 & 0.214 & $>$17   & $1.76\pm 0.15$ & 1\,TeV\\

%%%%%%%%%%%%%%%%%%%%%%%%%%%%%%%%%%%%%%%%%%%%%%%%%%%%%%%%

\section{Potential for VHE $\gamma$-ray observations} 
\label{section:fifth}

The individual SEDs of the 1BIGB sources can be exploited to estimate the detectability of these blazars at VHE $\gamma$ rays. To illustrate this possibility, we have selected five promising objects. The choice of sources was done considering only those objects with known redshift, in order to correctly estimate the absorption of VHE $\gamma$-rays due to the interaction with the extragalactic background light, EBL \citep{1992ApJ...390L..49S}. Among those, we considered only blazars with a hard {\it Fermi}-LAT photon spectral index ($\Gamma$\,<\,2.0). The selected sources are listed in Table~\ref{tab:sources_cta}, and ordered by redshift. The table also reports the synchrotron peak location, the {\it Fermi}-LAT slope obtained from our fit to the 1-100\,GeV points, and the cut-off energy of the intrinsic spectrum assumed in our extrapolations.

%This bias results from the fact that, in EHSPs, thanks to the shift
We would like to note that all selected sources are EHSPs, and mostly this bias result from the synchrotron peak shifting towards high energies, so that the host galaxy thermal component emerge at optical frequencies, allowing a precise redshift measurement.
%Since we have shown that the {\it Fermi}-LAT spectral slope does not discriminate between EHBLs and HSPs in our sample,

\begin{figure*}
\centering
\vspace{-0cm}
\subfloat[][1BIGB~J225147.5$-$320611 extrapolated with power-law and cut-off at 700 GeV.]
{\includegraphics[width=0.45\textwidth]{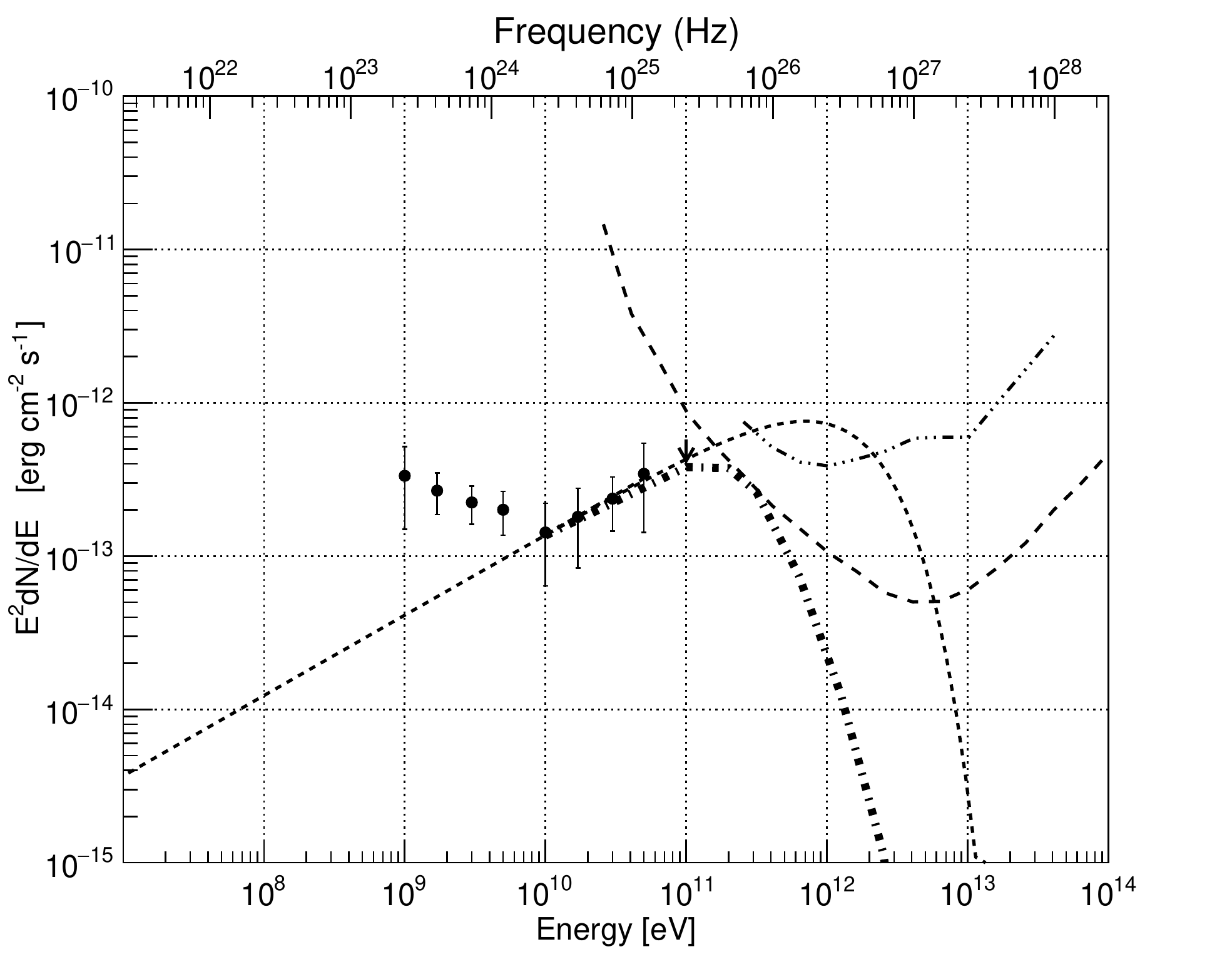} \label{CTA1}} \quad
\subfloat[][1BIGB~J223301.0+133601 extrapolated with power-law and cut-off at  1 TeV.]
{\includegraphics[width=0.45\textwidth]{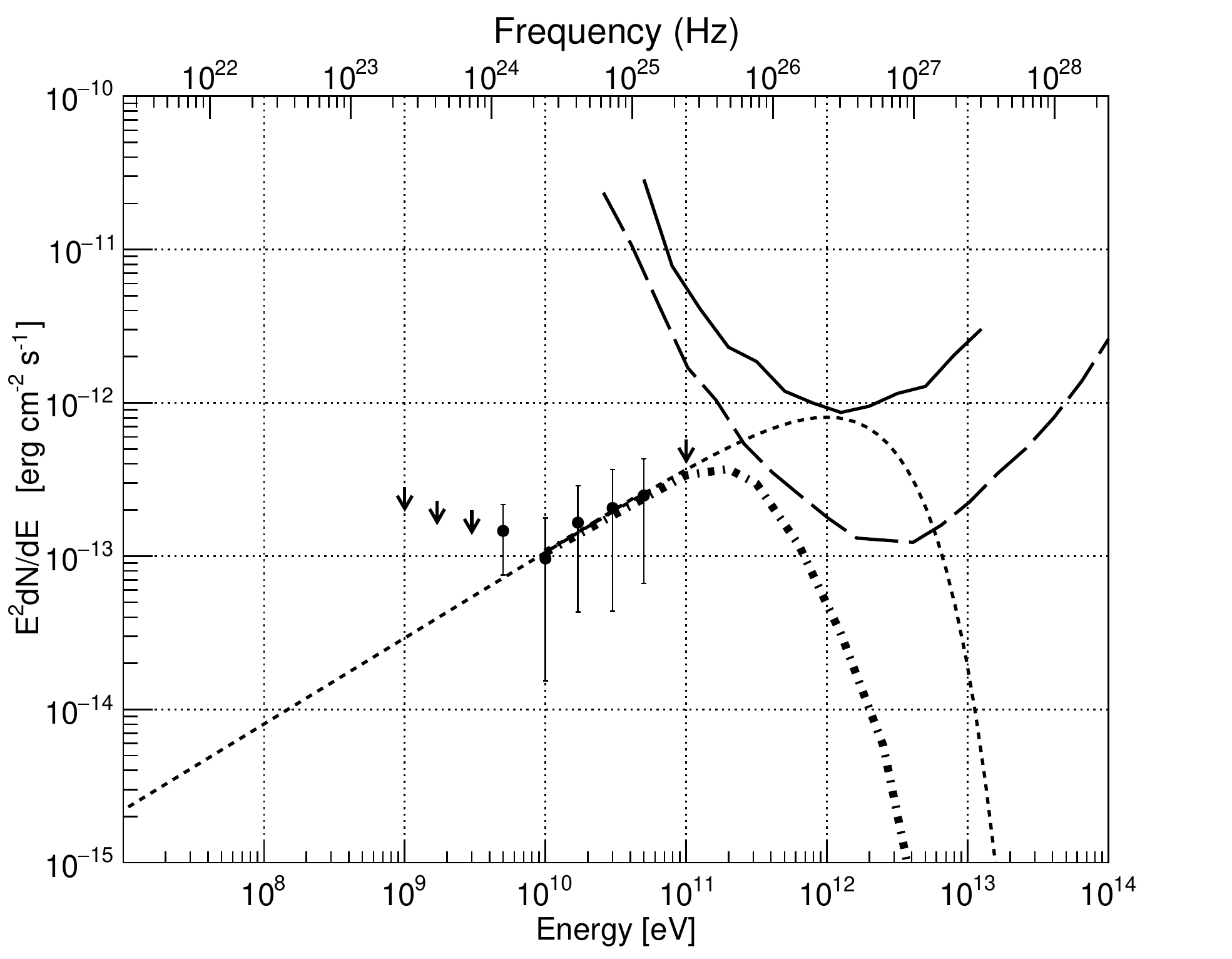} \label{CTA2}} \\
\subfloat[][1BIGB~J220155.8$-$170700 extrapolated with power-law and cut-off at 150 GeV.]
{\includegraphics[width=0.45\textwidth]{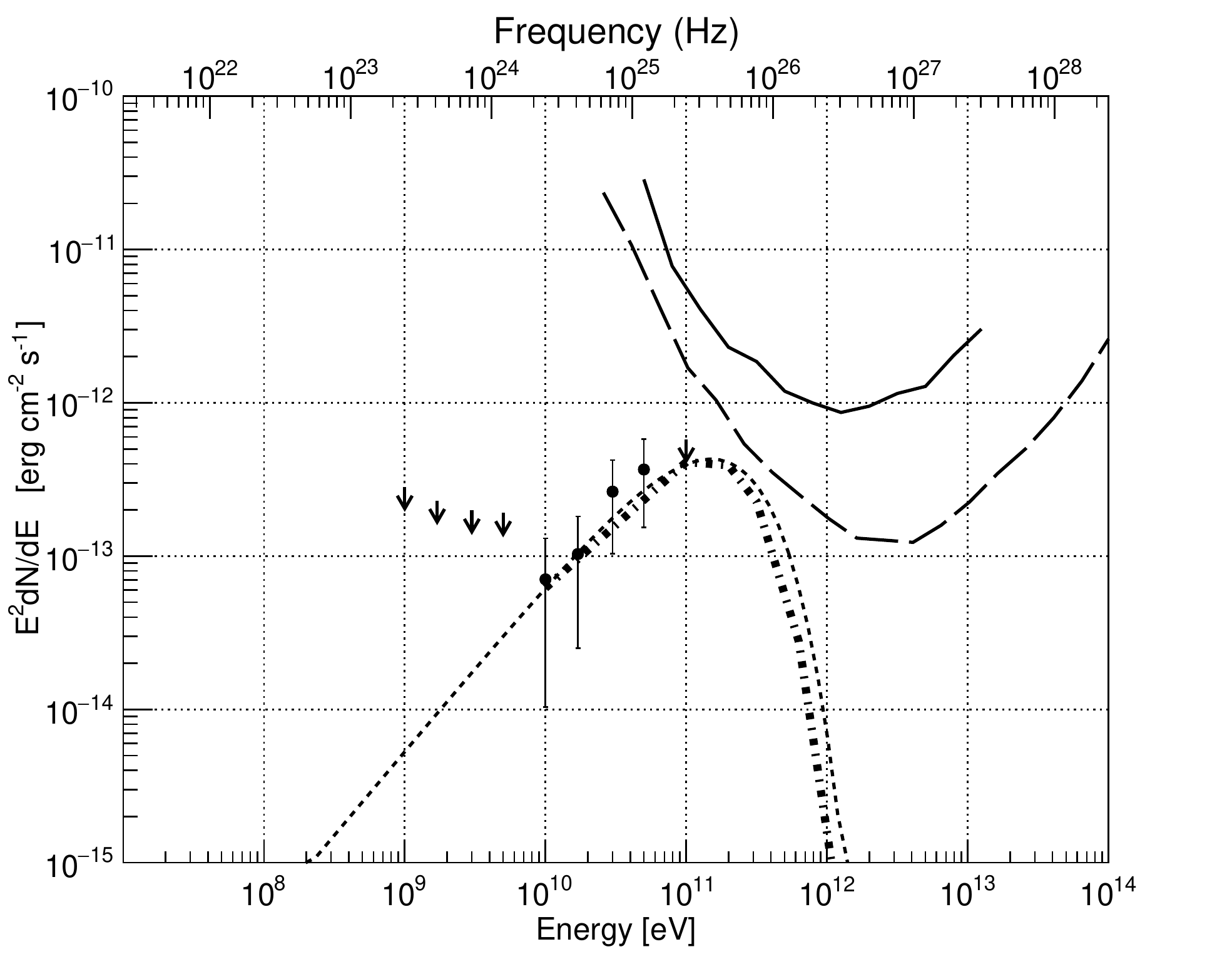} \label{CTA3}} \quad
\subfloat[][1BIGB~J151041.0+333503 extrapolated with power-law and cut-off at 300 GeV.]
{ \includegraphics[width=0.45\textwidth]{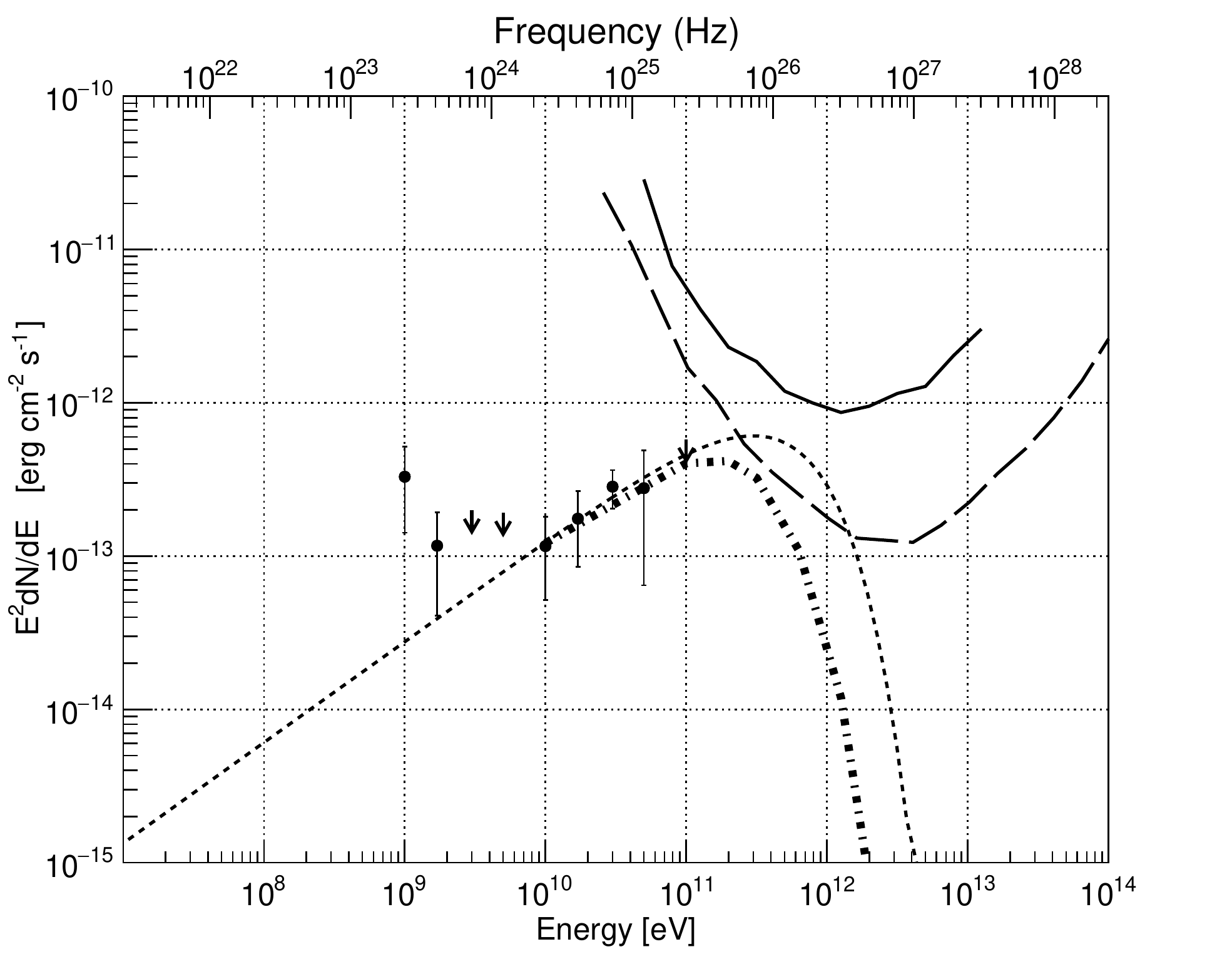} \label{CTA4}} \\
\subfloat[][1BIGB~J090802.2$-$095936 extrapolated with power-law and cut-off at 600 GeV.]
{\includegraphics[width=0.45\textwidth]{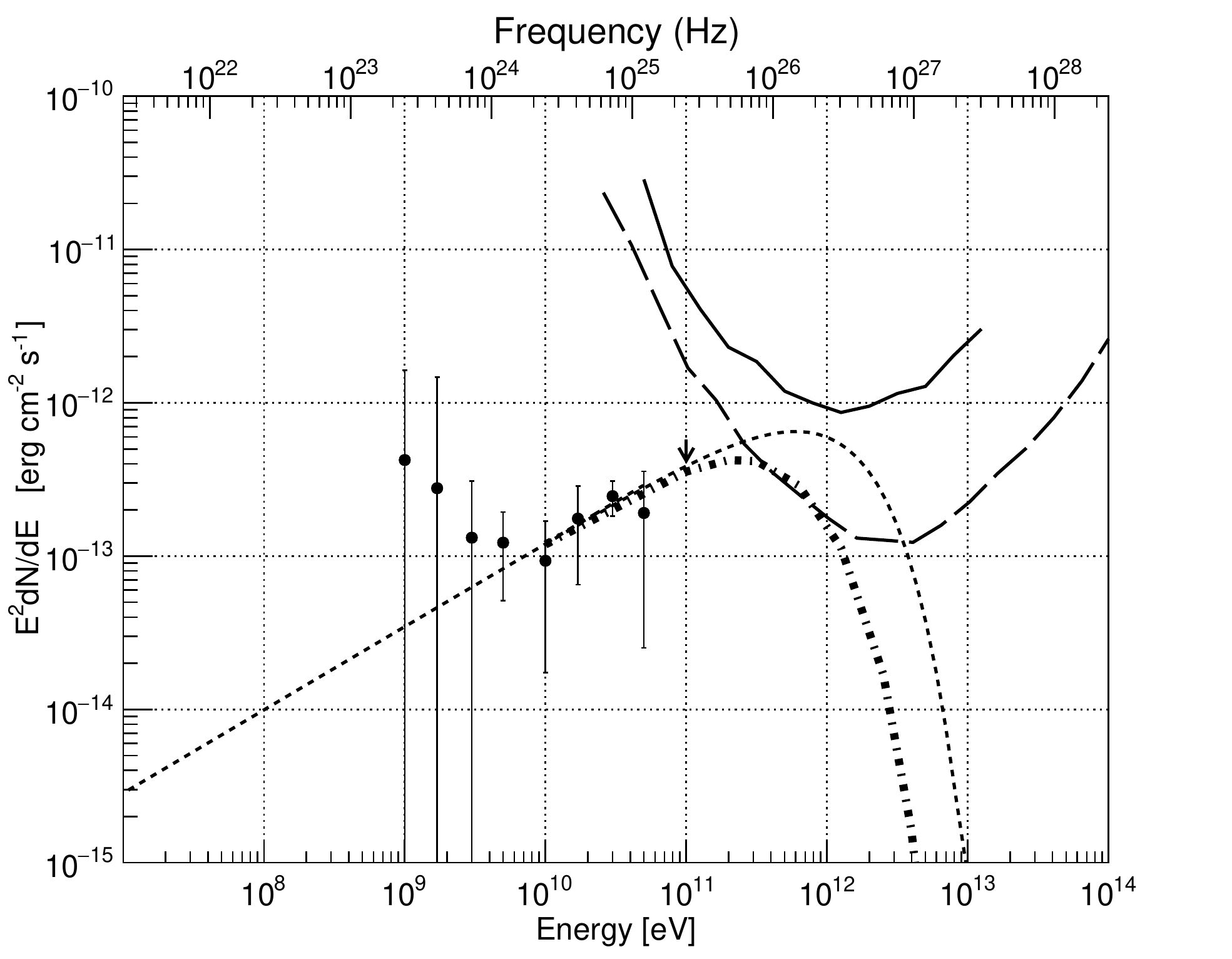} \label{CTA5}} \quad 
\subfloat[][Legend.]
{\adjustbox{raise={1.0cm}{\height}}{\includegraphics[width=0.4\textwidth]{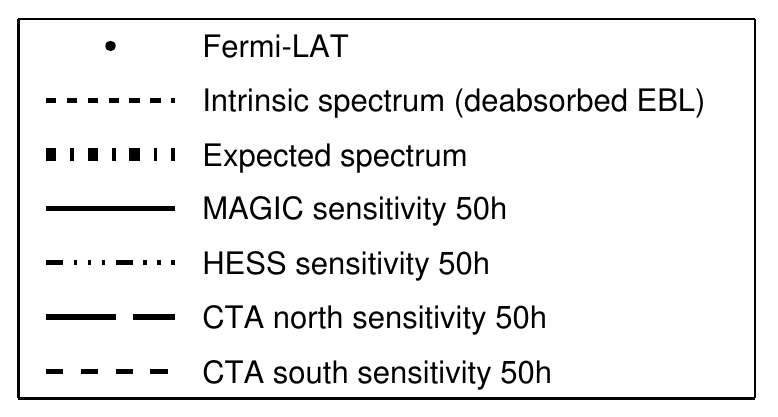}}}
\caption{Extrapolation of the $\gamma$-ray spectrum of five selected sources of the 1BIGB catalogue. The intrinsic spectrum assumed is a power-law with an exponential cut-off (dashed line). The resulting spectra once corrected for EBL absorption are displayed with dashed-dotted lines. CTA, MAGIC \citep{MAGICsens}, and H.E.S.S. \citep{HESSsens} sensitivities for 50\,h of observations are also reported in the plots.}
 \label{fig:CTA_extrapolations}
\end{figure*}

In \Cref{fig:CTA_extrapolations} we show the $\gamma$-ray SEDs of the selected sources up to the highest available energies. We extrapolate the \emph{Fermi}-LAT data points considering a power-law function with exponential cut-off. The cut-off energy for each source was estimated by taking into account all data points and upper limits within the 1-100\,GeV energy band. We considered absorption due to interaction with EBL based on $\tau_{(E,z)}$ values as predicted by \citet{2011MNRAS.410.2556D}. 

Depending on the location of each source, we have considered detectability with CTA-North and MAGIC, or CTA-South and H.E.S.S. . Each $\gamma$-ray SED is represented before (dashed line) and after (dotted-dashed line) EBL absorption. It is evident that the redshift is a very important parameter that can affect sensibly the detectability of blazar candidates. This is particularly true for EHSPs with an IC peak exceeding TeV energies, since in this case, already at relatively small redshifts such as those considered here, the absorption strongly affects the spectrum. The effect is evident in the two sources with the highest assumed cut-off energy, namely 1BIGB~J225147.5$-$320611 and 1BIGB~J223301.0+133601  , located at redshift 0.139 and 0.214. Thanks to the relatively high intrinsic emission, despite the EBL absorption severely affecting the observed flux, the extrapolations are still confidently within CTA detectability in both cases. %Both sources, may interestingly be in the reach of current-generation IACTs during high flux states.

For 1BIGB~J220155.8$-$170700 the upper limit of $4.19\cdot 10^{-13}$ erg/cm$^2$/s at 100 GeV derived from {\it Fermi}-LAT data severely constrains the location of the IC peak despite the negligible EBL absorption effect, and make this a difficult target even for the expected sensitivity of CTA-North. The last two sources, 1BIGB~J151041.0+333503 and 1BIGB~J090802.2-095936, present a spectrum compatible with an intrinsic IC peak location at hundreds GeV, and seems on the reach of CTA-North at least in the energy interval close to the SED peak.

All the extrapolated fluxes shown in \Cref{fig:CTA_extrapolations} are  well below the sensitivity of current generation of IACTs, indicating that a VHE $\gamma$-ray emission from these sources could be detected only in case of extremely bright states.

%The CTA visibility of these sources is strongly dependent on some parameters: redshift of the source, slope in the HE band, exponential cut-off value assumed for that source \dots 

%\begin{comment}
%\begin{figure}
%\centering
%\includegraphics[width=1.0\linewidth]{fig/1BIGB_J0304-0054.pdf}
%\caption{1BIGB J0304-0054 with CTA north/south sensitivity.}
%\label{CTA1}%
%\end{figure}

%   \begin{figure}
%   \centering
%   \includegraphics[width=1.0\linewidth]{fig/1BIGB_J0304.pdf}
%   \caption{1BIGB J0304-0054 with Magic Sensitivity.}
%              \label{CTA2}%
%    \end{figure}

%   \begin{figure}
%   \centering
%   \includegraphics[width=1.0\linewidth]{fig/1BIGB_J0442+6140.pdf}
%   \caption{1BIGB J0442+6140 with with CTA north/south Sensitivity.}
%              \label{CTA3}%
%    \end{figure}

%   \begin{figure}
%   \centering
%   \includegraphics[width=1.0\linewidth]{fig/LAST5POINTS_1BIGB_J0304-0054.pdf}
%   \caption{1BIGB J0304-0054 with with CTA north/south Sensitivity [Last five points - Luca].}
%              \label{CTA4}%
%    \end{figure}
%\end{comment}

%
%                                                One column figure
%----------------------------------------------------------- S_vib
%   \begin{figure}
%   \centering
%   %%%\includegraphics[width=3cm]{empty.eps}
%      \caption{Vibrational stability equation of state
%               $S_{\mathrm{vib}}(\lg e, \lg \rho)$.
%               $>0$ means vibrational stability.
%              }
%         \label{FigVibStab}
%   \end{figure}
%
%______________________________________________________________

\section{Conclusions}

%\item gamma-ray SEDs are at disposal
%\item population of promising TeV candidates for CTA
%\item Template representing a population of blazar that might be faint for Fermi-LAT (considering its sensitivity window) but still in reach for CTA observatory. 

We have presented for the first time the 1-100 GeV spectrum of a population of $\gamma$-ray faint EHSP sources, recently detected with {\it Fermi}-LAT in the 1BIGB catalogue. The results presented here are an update of the previous 1BIGB work, in the sense that we now include new PASS8 data integrated over 9.3 instead of 7.2 years of observations. This is the largest catalogue of extreme and weak blazars observed in $\gamma$ rays. All the data presented here is made available in VO tables via the Brazilian Science Data Center (BSDC) service (www.bsdc.cbpf.br), maintained at CBPF, and the BSDC Virtual Observatory (\url{vo.bsdc.icranet.org}). It is also accessible via the Open Universe Initiative Portal (www.openuniverse.asi.it) and through the SSDC SED Builder tool (\url{tools.asdc.asi.it}).

%\footnote{A templates with all the 148 SEDs will be available to be uploaded via SSDC SED builder.}.    

The 1BIGB sample discussed in this work consists of both HSP and EHSP blazars, whose spectral signatures in the {\it Fermi}-LAT band were shown to be similar among the two classes, and likely dependent on the source state. No spectral template for the classes was derivable, nevertheless we call attention to fact that the sample is limited in size and inhomogeneous.  

% With respect to expectations derived from the blazar sequence, we would like to note that nearly half of the sample has no measured redshift, as is usually the case for HSP blazars. 

%No spectral template for the classes was derivable, as one might expect possible from a description based on the blazar sequence. This may nevertheless be due to the fact that the sample is limited and inhomogeneous. 

The average observed spectral index in the 1-100 GeV band ranges between 1.3 and 2.8 indicating a predominantly hard spectrum. The average flux of the extreme HSP sample is relatively low, being composed of weak sources at the limit of the {\it Fermi}-LAT sensitivity. For these reasons we argue that EHSPs may escape {\it Fermi}-LAT detection (or be detected with low significance only), but still be prime candidates for observations with Cherenkov telescopes, due to the expected high VHE $\gamma$-ray emission. As we showed, it is expected for at least part of the objects, that the IC peak will fall well within the VHE band, above 100 GeV. Methods of selecting EHSPs not based exclusively on {\it Fermi}-LAT detections or extrapolations are therefore extremely interesting for completing the extragalactic population to be targeted by CTA.

In fact, we performed a detailed study of the potential for VHE $\gamma$-ray observations of the strongest 1BIGB sources with known redshift, indicating that they are well within the reach of future CTA capabilities. These analyses show that future VHE observations are in a good position to probe this poorly known, extreme class of HSP sources, expanding our understanding of the blazar phenomenon and bringing new elements to the framework of the blazar sequence scenario.

%\st{More interestingly, we identify that two of them, namely, 1BIGB J2251475-32061 and 1BIGB J2233010+13360, may be potential targets for current ground-based $\gamma$-ray observatories.}

%For WHSP catalogues \citep{1WHSP} and \cite{2WHSP}; For 1BIGB paper \cite{1BIGB}; For CTA concept  \cite{CTA} and CTA sensitivity \cite{CTA50h}.
%\cite{Neutrino-HSP} 
%We thank the State University of Campinas - Unicamp, IFGW Department of Physics for hosting the author. 

\section*{Acknowledgements}

During this work, BA was supported by S\~ao Paulo Research Foundation (FAPESP) with grant n. 2017/00517-4. UBdA acknowledges support from a FAPERJ Young Scientist Fellowship n. E10/2016-226465 and a CNPq Level 2 Fellowship n. 310827/2016-7. BF is supported by FAPERJ grants n. 202.687/2016 and 202.688/2016. We would like to thanks Prof. Marcelo M. Guzzo and Prof. Orlando L. G. Peres for the full support endorsing the author's partnership with FAPESP, and granting access to the Feynman Cluster from CCJDR Data Center at IFGW Unicamp, Campinas-Brazil. We also thank IcraNet, Prof. Remo Ruffini and Prof. Carlo Bianco for the cooperation granting access to Joshua Computer Cluster (Rome-Italy). The availability of computational resources was key to the development of our work. We make use of archival data and bibliographic information obtained from the NASA-IPAC Extragalactic Database (NED), data, and software facilities from the Space Science Data Center (SSDC) from the Italian Space Agency. The VO publication of our data (\url{vo.bsdc.icranet.org}) is made by the Brazilian Science Data Center (BSDC) service maintained at CBPF, Rio de Janeiro, and accessible through the United Nations Open Universe Initiative at http://www.openuniverse.asi.it. We thank Mr. Carlos Brandt for his support on this task. This research has made use of the CTA instrument response functions provided by the CTA Consortium and Observatory, see \url{http://www.cta-observatory.org/science/cta-performance/} (version prod3b-v1) for more details.
      
\bibliographystyle{mnras}
\bibliography{1BIGBSEDpaper}

\appendix
\section{The 1BIGB complete table with broadband fitting information}

%LONG-TABLE-1
\begin{table*}
\caption{Table showing the 148 1BIGB $\gamma$-ray signatures. The first three columns show respectively the 1BIGB source names, right ascension R.A. and declination Dec. in degrees (J2000). The fourth column shows the reported redshifts from literature \citep{bllacz2,pita2013,pks1424p240HighZ,PG1553,shaw2,masetti2013,bll,5BZcat},  flag ? is used for values reported as uncertain; lower limit are marked with ``$>$" \citep[all lower-limits shown here were derived in][]{1WHSP,2WHSP}, and sources with currently absent redshift were given 0. value. The $\gamma$-ray model parameters from Fermi Science Tools assume a power law to describe the spectrum within the studied energy range 0.3-500 GeV. The parameter $N_0$ (see eq. \ref{powerlaw}) is given in units of [ph/cm$^{2}$/s/MeV], and $\Gamma$ is the spectral photon index, which are direct outputs from the likelihood analysis over 7.2 years of Fermi-LAT data in the 0.3-500\,GeV band; those results consider the pivot energy fixed as E$_0$\,=\,1\,GeV. The column Flux gives the photon counts in units of ph/cm$^2$/s calculated by integrating eq. \ref{powerlaw} along the energy range 1-100\,GeV, column E-FLux corresponds to the energy flux in units of MeV/cm$^2$/s. For the columns Flux and E-flux, upper and lower case values represent positive and negative errors, respectively.}
\label{table:1BIGB}
 \begin{tabular}{llrcccccc}
\hline
1BIGB Source name  & R.A.(deg)  &  Dec.(deg)  &  z  &  $\Gamma$ & $N_0$ (10$^{-15}$)  & TS & Flux$_{1-100GeV}^{( \times 10^{-10})}$ & E-Flux$_{1-100GeV}^{( \times 10^{-13})}$ \\
 \hline
  1BIGBJ000949.6-431650  &   2.45708    &   -43.28056  &  $>$0.56    &  2.19$\pm$0.13  &   2.29$\pm$0.40   &   86.2 &  1.92$^{+0.63}_{-0.50}$   &  7.0$^{+2.3}_{-1.5}$ \\ 
  1BIGBJ001328.8+094929  &   3.37       &   9.825      &     0.      &  2.07$\pm$0.19  &   1.56$\pm$0.57   &   24.6 &  1.44$^{+0.95}_{-0.67}$   &  6.0$^{+3.5}_{-1.9}$ \\ 
  1BIGBJ001527.8+353638  &   3.86625    &   35.61083   &  $>$0.57    &  1.90$\pm$0.23  &   0.94$\pm$0.48   &   35.3 &  1.02$^{+0.99}_{-0.62}$   &  5.4$^{+4.7}_{-2.2}$ \\ 
  1BIGBJ002928.6+205332  &   7.36917    &   20.8925    &     0.      &  1.66$\pm$0.19  &   0.59$\pm$0.31   &   29.5 &  0.85$^{+0.87}_{-0.53}$   &  6.5$^{+5.3}_{-2.7}$ \\ 
  1BIGBJ004146.9-470136  &   10.44583   &   -47.02667  &     0.      &  1.79$\pm$0.27  &   0.33$\pm$0.21   &   10.8 &  0.40$^{+0.55}_{-0.30}$   &  2.5$^{+3.0}_{-1.2}$ \\ 
  1BIGBJ005816.6+172312  &   14.56958   &   17.38694   &     0.      &  1.79$\pm$0.24  &   0.73$\pm$0.44   &   24.0 &  0.89$^{+1.07}_{-0.62}$   &  5.6$^{+5.7}_{-2.5}$ \\ 
  1BIGBJ010250.8-200158  &   15.71208   &   -20.03278  &  $>$0.38    &  1.49$\pm$0.23  &   0.25$\pm$0.17   &   20.4 &  0.45$^{+0.68}_{-0.35}$   &  4.6$^{+5.2}_{-2.2}$ \\ 
  1BIGBJ011501.6-340027  &   18.75708   &   -34.0075   &     0.48    &  1.65$\pm$0.16  &   0.82$\pm$0.36   &   70.8 &  1.19$^{+0.98}_{-0.64}$   &  9.2$^{+6.1}_{-3.4}$ \\ 
  1BIGBJ012657.1+330730  &   21.73833   &   33.125     &     0.      &  2.19$\pm$0.29  &   1.38$\pm$0.57   &   22.2 &  1.15$^{+0.99}_{-0.61}$   &  4.2$^{+3.9}_{-1.6}$ \\ 
  1BIGBJ014040.8-075849  &   25.17      &   -7.98028   &  $>$0.49    &  1.77$\pm$0.14  &   1.12$\pm$0.37   &   45.0 &  1.41$^{+0.85}_{-0.61}$   &  9.1$^{+4.7}_{-2.9}$ \\ 
  1BIGBJ020106.1+003400  &   30.27542   &   0.56667    &     0.298   &  1.70$\pm$0.24  &   0.47$\pm$0.30   &   19.9 &  0.65$^{+0.84}_{-0.47}$   &  4.7$^{+5.0}_{-2.2}$ \\ 
  1BIGBJ020412.9-333339  &   31.05375   &   -33.56111  &     0.617   &  1.85$\pm$0.19  &   0.87$\pm$0.35   &   31.7 &  1.00$^{+0.78}_{-0.51}$   &  5.7$^{+4.1}_{-2.1}$ \\ 
  1BIGBJ021205.6-255757  &   33.02375   &   -25.96611  &     0.      &  1.99$\pm$0.14  &   1.67$\pm$0.41   &   58.9 &  1.67$^{+0.74}_{-0.56}$   &  7.8$^{+3.3}_{-2.1}$ \\ 
  1BIGBJ021216.8-022155  &   33.07      &   -2.36528   &     0.      &  1.92$\pm$0.19  &   1.16$\pm$0.44   &   31.8 &  1.23$^{+0.89}_{-0.60}$   &  6.3$^{+4.3}_{-2.3}$ \\ 
  1BIGBJ021631.9+231449  &   34.13333   &   23.24722   &     0.288   &  1.88$\pm$0.10  &   2.49$\pm$0.52   &  116.4 &  2.76$^{+0.98}_{-0.79}$   & 15.0$^{+4.7}_{-3.3}$ \\ 
  1BIGBJ022048.4-084250  &   35.20167   &   -8.71389   &  $>$0.43    &  1.88$\pm$0.20  &   1.00$\pm$0.43   &   34.1 &  1.11$^{+0.90}_{-0.59}$   &  6.0$^{+4.3}_{-2.2}$ \\ 
  1BIGBJ023340.9+065611  &   38.42042   &   6.93639    &     0.      &  2.03$\pm$0.14  &   3.23$\pm$0.90   &   98.0 &  3.08$^{+1.45}_{-1.12}$   & 13.6$^{+5.4}_{-3.5}$ \\ 
  1BIGBJ023430.5+804336  &   38.6275    &   80.72694   &     0.      &  1.55$\pm$0.14  &   0.37$\pm$0.16   &   32.4 &  0.63$^{+0.49}_{-0.34}$   &  5.8$^{+3.2}_{-1.9}$ \\ 
  1BIGBJ030103.7+344100  &   45.26542   &   34.68361   &     0.24    &  2.33$\pm$0.16  &   3.26$\pm$0.63   &   56.3 &  2.43$^{+0.87}_{-0.68}$   &  7.6$^{+2.7}_{-1.7}$ \\ 
  1BIGBJ030330.1+055429  &   45.87542   &   5.90833    &     0.196   &  1.50$\pm$0.23  &   0.33$\pm$0.26   &   24.5 &  0.60$^{+0.99}_{-0.51}$   &  6.0$^{+7.0}_{-3.0}$ \\ 
  1BIGBJ030433.9-005403  &   46.14125   &   -0.90111   &     0.511   &  1.69$\pm$0.16  &   0.79$\pm$0.35   &   38.1 &  1.09$^{+0.87}_{-0.59}$   &  7.9$^{+5.0}_{-2.8}$ \\ 
  1BIGBJ030544.1+403509  &   46.43375   &   40.58611   &     0.      &  1.88$\pm$0.25  &   0.77$\pm$0.46   &   19.0 &  0.85$^{+0.98}_{-0.58}$   &  4.6$^{+4.5}_{-2.0}$ \\ 
  1BIGBJ031103.1-440227  &   47.76333   &   -44.04111  &     0.      &  1.96$\pm$0.35  &   0.72$\pm$0.43   &   16.4 &  0.73$^{+1.03}_{-0.52}$   &  3.5$^{+5.5}_{-1.8}$ \\ 
  1BIGBJ031423.8+061955  &   48.59958   &   6.33222    &     0.62?   &  1.78$\pm$0.11  &   2.25$\pm$0.58   &   92.4 &  2.78$^{+1.22}_{-0.95}$   & 17.6$^{+6.4}_{-4.4}$ \\ 
  1BIGBJ032009.1-704533  &   50.03833   &   -70.75917  &     0.      &  1.77$\pm$0.18  &   0.63$\pm$0.26   &   32.1 &  0.79$^{+0.63}_{-0.41}$   &  5.1$^{+3.5}_{-1.9}$ \\ 
  1BIGBJ032037.9+112451  &   50.15833   &   11.41444   &     0.      &  2.36$\pm$0.24  &   3.12$\pm$0.99   &   24.8 &  2.27$^{+1.35}_{-0.95}$   &  6.9$^{+3.9}_{-2.1}$ \\ 
  1BIGBJ032056.2+042447  &   50.23458   &   4.41333    &     0.      &  2.70$\pm$0.22  &   3.03$\pm$0.61   &   30.1 &  1.78$^{+0.69}_{-0.52}$   &  4.1$^{+1.5}_{-0.9}$ \\ 
  1BIGBJ032647.2-340446  &   51.69708   &   -34.07972  &     0.      &  2.04$\pm$0.12  &   1.96$\pm$0.38   &   76.7 &  1.86$^{+0.64}_{-0.51}$   &  8.1$^{+2.7}_{-1.8}$ \\ 
  1BIGBJ032852.6-571605  &   52.21917   &   -57.26806  &     0.      &  1.55$\pm$0.18  &   0.36$\pm$0.19   &   36.3 &  0.61$^{+0.65}_{-0.39}$   &  5.6$^{+4.6}_{-2.4}$ \\ 
  1BIGBJ033623.7-034738  &   54.09875   &   -3.79389   &     0.162   &  1.44$\pm$0.19  &   0.36$\pm$0.23   &   41.6 &  0.71$^{+0.93}_{-0.52}$   &  7.8$^{+7.1}_{-3.5}$ \\ 
  1BIGBJ033831.9-570447  &   54.63333   &   -57.08     &     0.      &  1.89$\pm$0.27  &   0.43$\pm$0.25   &   10.6 &  0.48$^{+0.56}_{-0.32}$   &  2.5$^{+2.8}_{-1.1}$ \\ 
  1BIGBJ035856.1-305447  &   59.73375   &   -30.91306  &     0.65?   &  1.92$\pm$0.14  &   1.43$\pm$0.40   &   59.6 &  1.53$^{+0.76}_{-0.57}$   &  7.9$^{+3.6}_{-2.2}$ \\ 
  1BIGBJ041112.2-394143  &   62.80125   &   -39.69528  &  $>$0.7     &  1.74$\pm$0.29  &   0.23$\pm$0.19   &    6.7 &  0.31$^{+0.53}_{-0.26}$   &  2.1$^{+3.0}_{-1.1}$ \\ 
  1BIGBJ041238.3-392629  &   63.16      &   -39.44139  &     0.      &  1.91$\pm$0.30  &   0.34$\pm$0.25   &    6.7 &  0.37$^{+0.54}_{-0.29}$   &  1.9$^{+2.5}_{-9.5}$ \\ 
  1BIGBJ042900.1-323641  &   67.25042   &   -32.61139  &  $>$0.51    &  2.04$\pm$0.17  &   1.20$\pm$0.36   &   27.9 &  1.13$^{+0.63}_{-0.45}$   &  4.9$^{+2.6}_{-1.5}$ \\ 
  1BIGBJ043517.7-262121  &   68.82375   &   -26.35611  &     0.      &  2.60$\pm$0.28  &   1.64$\pm$0.43   &   20.2 &  1.02$^{+0.54}_{-0.38}$   &  2.5$^{+1.4}_{-0.7}$ \\ 
  1BIGBJ044127.4+150454  &   70.36417   &   15.08194   &     0.109   &  2.10$\pm$0.18  &   3.03$\pm$1.18   &   43.0 &  2.74$^{+1.81}_{-1.30}$   & 11.1$^{+6.0}_{-3.4}$ \\ 
  1BIGBJ044240.6+614039  &   70.66917   &   61.6775    &     0.      &  2.01$\pm$0.09  &   4.00$\pm$0.71   &  145.0 &  3.92$^{+1.12}_{-0.95}$   & 17.9$^{+4.3}_{-3.2}$ \\ 
  1BIGBJ044328.3-415156  &   70.86833   &   -41.86556  &  $>$0.39    &  1.96$\pm$0.15  &   1.58$\pm$0.44   &   66.8 &  1.63$^{+0.83}_{-0.61}$   &  8.0$^{+3.8}_{-2.3}$ \\ 
  1BIGBJ050335.3-111506  &   75.8975    &   -11.25167  &  $>$0.57    &  1.85$\pm$0.11  &   2.30$\pm$0.53   &  102.6 &  2.63$^{+1.03}_{-0.82}$   & 15.0$^{+5.1}_{-3.5}$ \\ 
  1BIGBJ050419.5-095631  &   76.08125   &   -9.94222   &  $>$0.46    &  2.14$\pm$0.23  &   1.29$\pm$0.54   &   16.6 &  1.12$^{+0.86}_{-0.58}$   &  4.3$^{+3.0}_{-1.5}$ \\ 
  1BIGBJ050601.6-382054  &   76.50667   &   -38.34861  &     0.182   &  1.95$\pm$0.11  &   2.06$\pm$0.40   &   96.5 &  2.13$^{+0.73}_{-0.59}$   & 10.5$^{+3.4}_{-2.4}$ \\ 
  1BIGBJ050727.1-334635  &   76.86333   &   -33.77639  &     0.      &  1.79$\pm$0.11  &   1.84$\pm$0.43   &  125.0 &  2.25$^{+0.93}_{-0.72}$   & 14.1$^{+5.0}_{-3.5}$ \\ 
  1BIGBJ053626.8-254748  &   84.11167   &   -25.79667  &     0.      &  2.06$\pm$0.16  &   1.70$\pm$0.44   &   47.9 &  1.58$^{+0.76}_{-0.56}$   &  6.7$^{+3.1}_{-1.9}$ \\ 
  1BIGBJ053645.2-255841  &   84.18875   &   -25.97806  &     0.      &  1.92$\pm$0.18  &   1.20$\pm$0.44   &   33.8 &  1.28$^{+0.88}_{-0.60}$   &  6.6$^{+4.2}_{-2.3}$ \\ 
  1BIGBJ055716.7-061706  &   89.32      &   -6.285     &     0.      &  1.95$\pm$0.15  &   2.51$\pm$0.86   &   52.2 &  2.59$^{+1.52}_{-1.12}$   & 12.7$^{+6.3}_{-3.8}$ \\ 
  1BIGBJ060714.2-251859  &   91.80958   &   -25.31639  &     0.275   &  1.94$\pm$0.14  &   1.54$\pm$0.40   &   51.9 &  1.61$^{+0.75}_{-0.57}$   &  8.1$^{+3.4}_{-2.2}$ \\ 
  1BIGBJ062149.6-341148  &   95.45667   &   -34.19694  &     0.529   &  2.53$\pm$0.12  &   4.77$\pm$0.59   &   93.3 &  3.09$^{+0.68}_{-0.58}$   &  8.1$^{+1.6}_{-1.2}$ \\ 
  1BIGBJ062626.2-171045  &   96.60917   &   -17.17944  &  $>$0.7     &  2.00$\pm$0.18  &   1.80$\pm$0.66   &   34.6 &  1.78$^{+1.17}_{-0.82}$   &  8.2$^{+4.9}_{-2.7}$ \\ 
  1BIGBJ063014.9-201236  &   97.5625    &   -20.21     &     0.      &  1.56$\pm$0.33  &   0.37$\pm$0.47   &   16.2 &  0.62$^{+1.79}_{-0.72}$   &  5.6$^{+11.0}_{-3.3}$ \\ 
  1BIGBJ065932.8-674350  &   104.88708  &   -67.73056  &     0.      &  1.63$\pm$0.17  &   0.55$\pm$0.29   &   39.4 &  0.82$^{+0.78}_{-0.50}$   &  6.5$^{+4.6}_{-2.5}$ \\ 
 \hline
 \end{tabular}
\end{table*}

\begin{table*}
\caption{Continued.}
\label{table:1BIGB}
 \begin{tabular}{llrcccccc}
\hline
1BIGB Source name  & R.A.(deg)  &  Dec.(deg)  &  z  &  $\Gamma$ & $N_0$ (10$^{-15}$)  & TS & Flux$_{1-100GeV}^{( \times 10^{-10})}$ & E-Flux$_{1-100GeV}^{( \times 10^{-13})}$ \\
 \hline
 1BIGBJ071745.0-552021  &   109.4375   &   -55.33944  &     0.      &  2.12$\pm$0.14  &   2.32$\pm$0.53   &   60.9 &  2.05$^{+0.83}_{-0.65}$   &  8.1$^{+3.0}_{-2.0}$ \\ 
  1BIGBJ073152.6+280432  &   112.96958  &   28.07583   &     0.248   &  1.90$\pm$0.18  &   1.06$\pm$0.39   &   29.9 &  1.15$^{+0.79}_{-0.54}$   &  6.1$^{+3.8}_{-2.1}$ \\ 
  1BIGBJ073329.5+351542  &   113.37292  &   35.26167   &     0.177   &  2.56$\pm$0.23  &   1.71$\pm$0.42   &   21.0 &  1.09$^{+0.51}_{-0.37}$   &  2.8$^{+1.2}_{-0.7}$ \\  
  1BIGBJ075936.1+132116  &   119.90042  &   13.35472   &     0.      &  1.75$\pm$0.09  &   2.12$\pm$0.44   &  152.6 &  2.71$^{+0.98}_{-0.79}$   & 17.9$^{+5.5}_{-4.0}$ \\ 
  1BIGBJ080015.4+561107  &   120.06458  &   56.18528   &     0.      &  1.92$\pm$0.12  &   1.54$\pm$0.34   &   78.9 &  1.65$^{+0.65}_{-0.51}$   &  8.5$^{+3.1}_{-2.1}$ \\ 
  1BIGBJ082904.7+175415  &   127.27     &   17.90417   &     0.089   &  2.25$\pm$0.10  &   4.47$\pm$0.55   &  136.8 &  3.55$^{+0.79}_{-0.67}$   & 12.1$^{+2.7}_{-2.0}$ \\ 
  1BIGBJ083724.5+145819  &   129.3525   &   14.97222   &     0.278   &  1.75$\pm$0.15  &   1.07$\pm$0.37   &   50.6 &  1.37$^{+0.87}_{-0.61}$   &  9.1$^{+4.9}_{-2.9}$ \\ 
  1BIGBJ085749.8+013530  &   134.4575   &   1.59167    &     0.281   &  2.57$\pm$0.23  &   1.81$\pm$0.48   &   18.6 &  1.15$^{+0.56}_{-0.42}$   &  2.9$^{+1.3}_{-0.7}$ \\  
  1BIGBJ090802.2-095936  &   137.00917  &   -9.99361   &     0.053   &  1.79$\pm$0.26  &   0.68$\pm$0.49   &   22.5 &  0.83$^{+1.17}_{-0.65}$   &  5.2$^{+5.8}_{-2.4}$ \\ 
  1BIGBJ090953.2+310602  &   137.47167  &   31.10083   &     0.272   &  1.81$\pm$0.21  &   0.62$\pm$0.32   &   22.3 &  0.75$^{+0.73}_{-0.45}$   &  4.5$^{+3.7}_{-1.8}$ \\ 
  1BIGBJ091322.3+813305  &   138.34292  &   81.55139   &     0.639?  &  1.40$\pm$0.15  &   0.20$\pm$0.10   &   41.5 &  0.42$^{+0.42}_{-0.26}$   &  4.9$^{+3.4}_{-1.9}$ \\ 
  1BIGBJ091651.8+523827  &   139.21625  &   52.64111   &     0.19    &  1.97$\pm$0.17  &   1.17$\pm$0.35   &   60.6 &  1.18$^{+0.67}_{-0.48}$   &  5.7$^{+3.1}_{-1.8}$ \\ 
  1BIGBJ093239.2+104234  &   143.16375  &   10.70972   &     0.361   &  2.00$\pm$0.14  &   1.95$\pm$0.49   &   58.5 &  1.92$^{+0.87}_{-0.66}$   &  8.9$^{+3.7}_{-2.4}$ \\ 
  1BIGBJ093430.1-172120  &   143.62542  &   -17.35583  &     0.      &  1.75$\pm$0.22  &   0.79$\pm$0.50   &   20.9 &  1.02$^{+1.22}_{-0.72}$   &  6.7$^{+6.4}_{-2.9}$ \\ 
  1BIGBJ095224.1+750212  &   148.10042  &   75.03694   &     0.181   &  1.43$\pm$0.15  &   0.29$\pm$0.14   &   62.4 &  0.57$^{+0.54}_{-0.34}$   &  6.3$^{+4.2}_{-2.4}$ \\ 
  1BIGBJ095507.9+355100  &   148.78292  &   35.85      &     0.834   &  2.00$\pm$0.25  &   0.82$\pm$0.38   &   24.1 &  0.80$^{+0.74}_{-0.46}$   &  3.7$^{+3.3}_{-1.5}$ \\ 
  1BIGBJ095628.2-095719  &   149.1175   &   -9.95528   &     0.      &  1.80$\pm$0.26  &   0.60$\pm$0.40   &   16.1 &  0.73$^{+0.98}_{-0.54}$   &  4.5$^{+5.2}_{-2.1}$ \\ 
  1BIGBJ095849.8+703959  &   149.7075   &   70.66639   &     0.      &  1.98$\pm$0.23  &   0.99$\pm$0.43   &   37.1 &  0.99$^{+0.84}_{-0.54}$   &  4.7$^{+3.7}_{-1.8}$ \\ 
  1BIGBJ102100.3+162554  &   155.25125  &   16.43167   &     0.556   &  2.16$\pm$0.23  &   1.48$\pm$0.52   &   27.9 &  1.27$^{+0.86}_{-0.58}$   &  4.8$^{+3.2}_{-1.6}$ \\ 
  1BIGBJ104303.7+005420  &   160.76583  &   0.90556    &     0.      &  1.75$\pm$0.14  &   1.30$\pm$0.43   &   56.5 &  1.67$^{+1.00}_{-0.72}$   & 11.0$^{+5.5}_{-3.4}$ \\ 
  1BIGBJ104857.6+500945  &   162.24     &   50.1625    &     0.402   &  2.37$\pm$0.15  &   2.01$\pm$0.35   &   53.1 &  1.46$^{+0.47}_{-0.38}$   &  4.4$^{+1.4}_{-0.9}$ \\ 
  1BIGBJ105534.3-012616  &   163.89292  &   -1.43778   &     0.      &  1.66$\pm$0.11  &   1.53$\pm$0.43   &   99.0 &  2.20$^{+1.08}_{-0.82}$   & 16.9$^{+6.6}_{-4.5}$ \\ 
  1BIGBJ111717.5+000633  &   169.32292  &   0.10917    &     0.451   &  1.92$\pm$0.15  &   1.87$\pm$0.53   &   57.4 &  2.00$^{+1.04}_{-0.76}$   & 10.3$^{+4.9}_{-3.0}$ \\ 
  1BIGBJ112317.9-323217  &   170.825    &   -32.53833  &     0.      &  2.06$\pm$0.22  &   1.25$\pm$0.48   &   22.8 &  1.16$^{+0.85}_{-0.57}$   &  4.9$^{+3.5}_{-1.7}$ \\ 
  1BIGBJ112611.8-203723  &   171.54958  &   -20.62333  &     0.      &  2.12$\pm$0.27  &   1.15$\pm$0.59   &   11.6 &  1.02$^{+0.99}_{-0.62}$   &  4.0$^{+3.5}_{-1.6}$ \\ 
  1BIGBJ113046.0-313807  &   172.69208  &   -31.63528  &     0.151   &  1.29$\pm$0.20  &   0.13$\pm$0.10   &   24.8 &  0.33$^{+0.56}_{-0.28}$   &  4.7$^{+5.0}_{-2.3}$ \\ 
  1BIGBJ113105.2-094405  &   172.77167  &   -9.735     &     0.      &  1.66$\pm$0.14  &   0.99$\pm$0.35   &   64.1 &  1.43$^{+0.92}_{-0.65}$   & 10.9$^{+5.7}_{-3.5}$ \\ 
  1BIGBJ113444.6-172900  &   173.68625  &   -17.48361  &     0.571   &  1.68$\pm$0.17  &   0.79$\pm$0.35   &   36.9 &  1.10$^{+0.91}_{-0.60}$   &  8.2$^{+5.4}_{-3.0}$ \\ 
  1BIGBJ113755.6-171041  &   174.48167  &   -17.17833  &     0.6     &  1.71$\pm$0.10  &   1.90$\pm$0.43   &  126.9 &  2.55$^{+1.00}_{-0.79}$   & 18.0$^{+6.0}_{-4.2}$ \\ 
  1BIGBJ121158.6+224233  &   182.99417  &   22.70917   &     0.45    &  1.83$\pm$0.17  &   0.89$\pm$0.34   &   30.2 &  1.04$^{+0.74}_{-0.50}$   &  6.0$^{+3.8}_{-2.1}$ \\ 
  1BIGBJ121510.9+073203  &   183.79542  &   7.53444    &     0.137   &  1.64$\pm$0.11  &   1.26$\pm$0.37   &  105.5 &  1.87$^{+0.97}_{-0.72}$   & 14.9$^{+6.2}_{-4.1}$ \\ 
  1BIGBJ121603.1-024304  &   184.01333  &   -2.71778   &     0.169   &  2.22$\pm$0.16  &   2.81$\pm$0.67   &   56.3 &  2.28$^{+0.97}_{-0.74}$   &  8.0$^{+3.2}_{-2.0}$ \\ 
  1BIGBJ124141.4+344029  &   190.4225   &   34.675     &  $>$0.7     &  1.88$\pm$0.15  &   1.19$\pm$0.37   &   46.8 &  1.33$^{+0.76}_{-0.55}$   &  7.3$^{+3.8}_{-2.3}$ \\ 
  1BIGBJ125015.4+315559  &   192.56458  &   31.93306   &     0.      &  1.82$\pm$0.31  &   0.55$\pm$0.39   &   16.8 &  0.65$^{+1.02}_{-0.51}$   &  3.9$^{+5.7}_{-2.0}$ \\ 
  1BIGBJ125341.2-393159  &   193.42167  &   -39.53306  &     0.179   &  1.89$\pm$0.17  &   1.29$\pm$0.51   &   39.8 &  1.42$^{+1.00}_{-0.69}$   &  7.6$^{+4.6}_{-2.6}$ \\ 
  1BIGBJ125847.9-044744  &   194.7      &   -4.79583   &     0.586?  &  1.93$\pm$0.11  &   2.50$\pm$0.62   &   61.2 &  2.64$^{+1.10}_{-0.87}$   & 13.5$^{+4.7}_{-3.2}$ \\ 
  1BIGBJ130145.6+405623  &   195.44     &   40.94      &     0.652   &  2.18$\pm$0.15  &   2.04$\pm$0.40   &   71.5 &  1.71$^{+0.63}_{-0.49}$   &  6.3$^{+2.3}_{-1.5}$ \\ 
  1BIGBJ130713.3-034430  &   196.80542  &   -3.74194   &     0.      &  2.05$\pm$0.26  &   1.14$\pm$0.57   &   19.0 &  1.08$^{+1.03}_{-0.64}$   &  4.7$^{+4.2}_{-1.9}$ \\ 
  1BIGBJ132541.8-022809  &   201.42417  &   -2.46944   &     0.8?    &  1.93$\pm$0.18  &   1.13$\pm$0.42   &   27.3 &  1.19$^{+0.80}_{-0.56}$   &  6.0$^{+3.6}_{-2.0}$ \\ 
  1BIGBJ132617.7+122957  &   201.57375  &   12.49944   &     0.204   &  2.08$\pm$0.20  &   1.57$\pm$0.50   &   32.2 &  1.44$^{+0.87}_{-0.61}$   &  6.0$^{+3.6}_{-1.9}$ \\ 
  1BIGBJ132833.4+114520  &   202.13958  &   11.75556   &     0.811   &  1.92$\pm$0.17  &   1.31$\pm$0.46   &   37.5 &  1.40$^{+0.90}_{-0.63}$   &  7.1$^{+4.1}_{-2.3}$ \\ 
  1BIGBJ133612.1+231958  &   204.05042  &   23.33278   &     0.267   &  1.97$\pm$0.15  &   1.66$\pm$0.42   &   62.5 &  1.68$^{+0.79}_{-0.59}$   &  8.0$^{+3.6}_{-2.3}$ \\ 
  1BIGBJ135328.0+560056  &   208.36667  &   56.01556   &     0.404   &  2.25$\pm$0.15  &   2.13$\pm$0.42   &   71.0 &  1.69$^{+0.61}_{-0.48}$   &  5.7$^{+2.0}_{-1.3}$ \\ 
  1BIGBJ140629.9-393508  &   211.625    &   -39.58583  &     0.37    &  1.54$\pm$0.16  &   0.56$\pm$0.27   &   42.7 &  0.96$^{+0.87}_{-0.56}$   &  9.0$^{+5.9}_{-3.3}$ \\ 
  1BIGBJ143342.7-730437  &   218.42792  &   -73.07722  &     0.      &  1.33$\pm$0.20  &   0.15$\pm$0.11   &   24.6 &  0.36$^{+0.55}_{-0.29}$   &  4.7$^{+4.6}_{-2.2}$ \\ 
  1BIGBJ143825.4+120418  &   219.60625  &   12.07167   &     0.      &  2.11$\pm$0.21  &   1.35$\pm$0.49   &   25.2 &  1.20$^{+0.82}_{-0.56}$   &  4.8$^{+3.1}_{-1.6}$ \\ 
  1BIGBJ144236.4-462300  &   220.65167  &   -46.38361  &     0.103   &  1.92$\pm$0.10  &   3.14$\pm$0.64   &  107.7 &  3.33$^{+1.14}_{-0.93}$   & 17.1$^{+5.0}_{-3.6}$ \\ 
  1BIGBJ145543.6-760051  &   223.93167  &   -76.01444  &     0.      &  1.61$\pm$0.12  &   0.63$\pm$0.22   &   43.7 &  0.96$^{+0.59}_{-0.43}$   &  7.9$^{+3.6}_{-2.3}$ \\ 
  1BIGBJ145603.5+504825  &   224.015    &   50.80722   &  $>$0.49    &  2.09$\pm$0.24  &   1.23$\pm$0.55   &   27.2 &  1.11$^{+0.92}_{-0.60}$   &  4.5$^{+3.4}_{-1.6}$ \\ 
  1BIGBJ150637.0-054004  &   226.65458  &   -5.66778   &     0.518   &  1.72$\pm$0.17  &   0.94$\pm$0.41   &   35.2 &  1.25$^{+1.00}_{-0.67}$   &  8.7$^{+5.5}_{-3.1}$ \\ 
  1BIGBJ151041.0+333503  &   227.67125  &   33.58444   &     0.114   &  1.59$\pm$0.30  &   0.27$\pm$0.24   &   15.2 &  0.43$^{+0.91}_{-0.40}$   &  3.7$^{+6.4}_{-2.1}$ \\ 
  1BIGBJ151136.8-165326  &   227.90375  &   -16.89056  &  $>$0.56    &  2.15$\pm$0.35  &   1.22$\pm$0.74   &   13.1 &  1.06$^{+1.36}_{-0.74}$   &  4.0$^{+5.3}_{-1.8}$ \\ 
  1BIGBJ151618.7-152344  &   229.07792  &   -15.39556  &  $>$0.54    &  1.98$\pm$0.18  &   1.50$\pm$0.58   &   27.2 &  1.51$^{+1.02}_{-0.72}$   &  7.1$^{+4.1}_{-2.3}$ \\ 
  1BIGBJ151826.5+075222  &   229.61083  &   7.87278    &     0.41    &  1.74$\pm$0.20  &   0.66$\pm$0.33   &   24.6 &  0.85$^{+0.81}_{-0.51}$   &  5.7$^{+4.5}_{-2.3}$ \\ 
  1BIGBJ151845.7+061355  &   229.69042  &   6.23222    &     0.102   &  1.96$\pm$0.21  &   1.36$\pm$0.56   &   31.8 &  1.39$^{+1.09}_{-0.72}$   &  6.8$^{+5.0}_{-2.5}$ \\ 
  1BIGBJ152646.6-153025  &   231.69417  &   -15.50722  &  $>$0.43    &  2.04$\pm$0.13  &   2.61$\pm$0.64   &   58.4 &  2.47$^{+1.06}_{-0.82}$   & 10.8$^{+4.1}_{-2.7}$ \\ 
  1BIGBJ152913.5+381216  &   232.30625  &   38.20472   &  $>$0.59    &  2.04$\pm$0.19  &   1.22$\pm$0.39   &   34.6 &  1.16$^{+0.69}_{-0.49}$   &  5.1$^{+2.9}_{-1.6}$ \\ 
  1BIGBJ154202.9-291509  &   235.5125   &   -29.2525   &     0.      &  1.78$\pm$0.08  &   2.62$\pm$0.50   &  143.2 &  3.27$^{+1.04}_{-0.86}$   & 20.8$^{+5.4}_{-4.1}$ \\ 
  1BIGBJ154625.0-285723  &   236.60417  &   -28.95639  &  $>$0.6     &  1.73$\pm$0.18  &   0.67$\pm$0.36   &   18.7 &  0.88$^{+0.84}_{-0.54}$   &  6.0$^{+4.3}_{-2.3}$ \\ 
  1BIGBJ155053.2-082245  &   237.72167  &   -8.37944   &     0.      &  1.87$\pm$0.14  &   1.96$\pm$0.63   &   64.9 &  2.20$^{+1.24}_{-0.91}$   & 12.2$^{+5.8}_{-3.6}$ \\ 
 \hline
 \end{tabular}
\end{table*}

\begin{table*}
\caption{Continued.}
\label{table:1BIGB}
 \begin{tabular}{llrcccccc}
\hline
1BIGB Source name  & R.A.(deg)  &  Dec.(deg)  &  z  &  $\Gamma$ & $N_0$ (10$^{-15}$)  & TS & Flux$_{1-100GeV}^{( \times 10^{-10})}$ & E-Flux$_{1-100GeV}^{( \times 10^{-13})}$ \\
 \hline
  1BIGBJ155432.5-121324  &   238.63542  &   -12.22361  &     0.      &  1.69$\pm$0.13  &   1.04$\pm$0.39   &   34.3 &  1.45$^{+0.94}_{-0.68}$   & 10.6$^{+5.3}_{-3.3}$ \\ 
  1BIGBJ160218.0+305108  &   240.575    &   30.8525    &  $>$0.47    &  2.44$\pm$0.27  &   1.83$\pm$0.47   &   33.8 &  1.27$^{+0.70}_{-0.48}$   &  3.6$^{+2.3}_{-1.1}$ \\ 
  1BIGBJ160519.0+542058  &   241.32917  &   54.34972   &     0.212   &  1.88$\pm$0.13  &   1.13$\pm$0.29   &   64.0 &  1.25$^{+0.59}_{-0.45}$   &  6.8$^{+3.0}_{-1.9}$ \\ 
  1BIGBJ160618.4+134532  &   241.57667  &   13.75889   &     0.29    &  2.05$\pm$0.21  &   1.76$\pm$0.64   &   33.8 &  1.65$^{+1.14}_{-0.77}$   &  7.1$^{+4.7}_{-2.5}$ \\ 
  1BIGBJ161327.1-190835  &   243.36292  &   -19.14333  &     0.      &  2.14$\pm$0.15  &   3.15$\pm$0.82   &   34.1 &  2.73$^{+1.25}_{-0.95}$   & 10.5$^{+4.3}_{-2.7}$ \\ 
  1BIGBJ162115.1-003140  &   245.31333  &   -0.52778   &  $>$0.52    &  1.80$\pm$0.21  &   0.77$\pm$0.41   &   14.4 &  0.93$^{+0.93}_{-0.58}$   &  5.6$^{+4.7}_{-2.3}$ \\ 
  1BIGBJ162330.4+085724  &   245.87708  &   8.95667    &     0.533   &  1.91$\pm$0.20  &   1.23$\pm$0.52   &   35.5 &  1.33$^{+1.05}_{-0.69}$   &  7.0$^{+4.9}_{-2.5}$ \\ 
  1BIGBJ162646.0+630047  &   246.69167  &   63.01333   &     0.      &  1.91$\pm$0.17  &   1.00$\pm$0.33   &   51.6 &  1.07$^{+0.65}_{-0.46}$   &  5.5$^{+3.1}_{-1.8}$ \\ 
  1BIGBJ164220.2+221143  &   250.58458  &   22.19528   &     0.592   &  2.19$\pm$0.19  &   1.76$\pm$0.55   &   32.1 &  1.46$^{+0.83}_{-0.60}$   &  5.3$^{+2.8}_{-1.6}$ \\ 
  1BIGBJ164419.9+454644  &   251.08333  &   45.77889   &     0.225   &  1.65$\pm$0.16  &   0.48$\pm$0.20   &   39.1 &  0.70$^{+0.56}_{-0.37}$   &  5.4$^{+3.5}_{-2.0}$ \\ 
  1BIGBJ165517.8-224045  &   253.82458  &   -22.67917  &     0.      &  1.99$\pm$0.21  &   1.64$\pm$0.75   &   17.4 &  1.64$^{+1.38}_{-0.91}$   &  7.7$^{+5.6}_{-2.8}$ \\ 
  1BIGBJ171108.5+024403  &   257.78583  &   2.73444    &     0.      &  1.83$\pm$0.18  &   0.93$\pm$0.42   &   18.3 &  1.09$^{+0.89}_{-0.60}$   &  6.4$^{+4.2}_{-2.3}$ \\ 
  1BIGBJ174419.7+185218  &   266.0825   &   18.87167   &     0.      &  1.58$\pm$0.21  &   0.41$\pm$0.26   &   15.5 &  0.65$^{+0.85}_{-0.48}$   &  5.7$^{+5.6}_{-2.6}$ \\ 
  1BIGBJ174702.5+493800  &   266.76042  &   49.63361   &     0.46?   &  2.27$\pm$0.22  &   1.74$\pm$0.49   &   37.3 &  1.37$^{+0.75}_{-0.53}$   &  4.5$^{+2.6}_{-1.4}$ \\ 
  1BIGBJ184822.4+653656  &   282.09375  &   65.61583   &     0.364   &  1.59$\pm$0.15  &   0.50$\pm$0.21   &   52.8 &  0.79$^{+0.63}_{-0.42}$   &  6.8$^{+4.2}_{-2.4}$ \\ 
  1BIGBJ185023.9+263153  &   282.6      &   26.53139   &     0.      &  1.63$\pm$0.13  &   1.20$\pm$0.45   &   88.6 &  1.80$^{+1.18}_{-0.85}$   & 14.5$^{+7.1}_{-4.5}$ \\ 
  1BIGBJ185813.3+432451  &   284.55583  &   43.41417   &     0.      &  2.12$\pm$0.16  &   1.89$\pm$0.48   &   43.3 &  1.67$^{+0.77}_{-0.58}$   &  6.6$^{+2.9}_{-1.8}$ \\ 
  1BIGBJ193412.7-241919  &   293.55292  &   -24.32222  &     0.      &  1.60$\pm$0.12  &   0.88$\pm$0.30   &   53.5 &  1.36$^{+0.81}_{-0.59}$   & 11.5$^{+5.2}_{-3.4}$ \\ 
  1BIGBJ194356.2+211821  &   295.98417  &   21.30611   &     0.      &  1.44$\pm$0.08  &   2.01$\pm$0.57   &  194.7 &  3.93$^{+1.84}_{-1.44}$   & 42.9$^{+13.2}_{-9.8}$ \\   
  1BIGBJ200204.0-573644  &   300.51708  &   -57.6125   &     0.      &  2.08$\pm$0.10  &   2.93$\pm$0.49   &  105.5 &  2.69$^{+0.78}_{-0.64}$   & 11.2$^{+3.0}_{-2.2}$ \\ 
  1BIGBJ201200.9-771219  &   303.00375  &   -77.20528  &     0.      &  2.10$\pm$0.30  &   0.78$\pm$0.52   &    9.8 &  0.71$^{+0.89}_{-0.52}$   &  2.8$^{+3.0}_{-1.2}$ \\ 
  1BIGBJ205242.4+081040  &   313.17708  &   8.17778    &     0.      &  1.51$\pm$0.23  &   0.25$\pm$0.20   &   15.3 &  0.45$^{+0.73}_{-0.37}$   &  4.3$^{+5.1}_{-2.2}$ \\ 
  1BIGBJ214533.3-043438  &   326.38875  &   -4.5775    &     0.069   &  2.64$\pm$0.31  &   1.34$\pm$0.47   &    9.7 &  0.82$^{+0.55}_{-0.37}$   &  1.9$^{+1.2}_{-0.6}$ \\ 
  1BIGBJ215214.0-120540  &   328.05875  &   -12.09472  &     0.121   &  2.26$\pm$0.14  &   3.77$\pm$0.76   &   81.3 &  2.98$^{+1.06}_{-0.84}$   & 10.0$^{+3.3}_{-2.2}$ \\ 
  1BIGBJ220107.3-590639  &   330.28042  &   -59.11111  &     0.      &  1.93$\pm$0.18  &   0.78$\pm$0.28   &   21.3 &  0.82$^{+0.55}_{-0.38}$   &  4.1$^{+2.4}_{-1.4}$ \\ 
  1BIGBJ220155.8-170700  &   330.4825   &   -17.11667  &     0.169   &  1.32$\pm$0.37  &   0.16$\pm$0.22   &   33.5 &  0.38$^{+1.63}_{-0.47}$   &  5.1$^{+14.1}_{-3.5}$ \\  
  1BIGBJ221029.5+362159  &   332.62333  &   36.36639   &     0.      &  2.17$\pm$0.32  &   1.19$\pm$0.75   &   13.3 &  1.01$^{+1.23}_{-0.72}$   &  3.7$^{+4.1}_{-1.6}$ \\ 
  1BIGBJ221108.2-000302  &   332.78458  &   -0.05056   &     0.326   &  1.84$\pm$0.14  &   1.43$\pm$0.47   &   48.7 &  1.67$^{+0.98}_{-0.71}$   &  9.7$^{+4.8}_{-2.9}$ \\ 
  1BIGBJ223301.0+133601  &   338.25458  &   13.60028   &     0.214   &  1.51$\pm$0.24  &   0.24$\pm$0.19   &   15.4 &  0.43$^{+0.73}_{-0.37}$   &  4.2$^{+5.1}_{-2.1}$ \\ 
  1BIGBJ223626.2+370713  &   339.10958  &   37.12028   &     0.      &  1.85$\pm$0.20  &   0.98$\pm$0.48   &   29.1 &  1.12$^{+1.01}_{-0.66}$   &  6.3$^{+4.8}_{-2.4}$ \\ 
  1BIGBJ224910.6-130002  &   342.29458  &   -13.00056  &  $>$0.5     &  2.33$\pm$0.01  &  87.01$\pm$1.43   &11240.5 &  64.8$^{+1.93}_{-1.88}$   &202.8$^{+6.2}_{-6.0}$ \\ 
  1BIGBJ225147.5-320611  &   342.94792  &   -32.10333  &     0.246   &  2.07$\pm$0.19  &   1.60$\pm$0.51   &   48.6 &  1.48$^{+0.90}_{-0.63}$   &  6.2$^{+3.7}_{-2.0}$ \\ 
  1BIGBJ225613.3-330338  &   344.05542  &   -33.06056  &     0.243   &  2.56$\pm$0.21  &   2.23$\pm$0.42   &   39.7 &  1.42$^{+0.54}_{-0.41}$   &  3.6$^{+1.4}_{-0.8}$ \\  
  1BIGBJ230634.9-110347  &   346.64583  &   -11.06333  &     0.      &  1.69$\pm$0.21  &   0.62$\pm$0.34   &   28.5 &  0.86$^{+0.94}_{-0.56}$   &  6.3$^{+5.7}_{-2.7}$ \\ 
  1BIGBJ232039.7-630918  &   350.16583  &   -63.155    &     0.2     &  1.84$\pm$0.16  &   0.84$\pm$0.30   &   43.6 &  0.98$^{+0.64}_{-0.45}$   &  5.7$^{+3.2}_{-1.8}$ \\ 
  1BIGBJ233112.8-030129  &   352.80375  &   -3.025     &     0.      &  2.05$\pm$0.15  &   1.95$\pm$0.54   &   40.8 &  1.84$^{+0.90}_{-0.67}$   &  7.9$^{+3.5}_{-2.2}$ \\ 
  1BIGBJ235320.9-145856  &   358.3375   &   -14.9825   &     0.      &  1.83$\pm$0.28  &   0.55$\pm$0.37   &   13.3 &  0.65$^{+0.92}_{-0.49}$   &  3.8$^{+4.9}_{-1.9}$ \\ 
 \hline
 \end{tabular}
\end{table*}

%Redshift references. \citep{bllacz2,pita2013,pks1424p240HighZ,PG1553,shaw2,masetti2013,bll,5BZcat},

%\include{1BIGB-9yrs-long-table}

%%%%%%%%%%%%%%%%%%%%%%%%%%%%%%%%%%%%%%%%%%%%%%%%%%



%\bibliographystyle{mnras}

%\begin{thebibliography}{99}
%\end{thebibliography}

%%%%%%%%%%%%%%%%%%%%%%%%%%%%%%%%%%%%%%%%%%%%%%%%%%

%%%%%%%%%%%%%%%%% APPENDICES %%%%%%%%%%%%%%%%%%%%%

%\appendix
%\newpage
%\section{Updating the 1BIGB broadband fitting at 0.3-500\,GeV: The complete table}
%\label{appendix1}
%\include{1BIGB-9yrs-long-table}


%%%%%%%%%%%%%%%%%%%%%%%%%%%%%%%%%%%%%%%%%%%%%%%%%%

% Don't change these lines
\bsp	% typesetting comment
\label{lastpage}
\end{document}